\documentclass[aps,preprintnumbers,twocolumn,amsmath,nofootinbib,amssymb]{revtex4}
\usepackage{graphicx,color,dcolumn,booktabs,bm}
\usepackage{longtable,lscape}
\usepackage{txfonts}
\usepackage{overpic}
\usepackage{epstopdf}
\usepackage{indentfirst}
\usepackage{feynmf}   
\usepackage{slashed}  
\usepackage{cases}
\usepackage{color}
\usepackage{float}
\usepackage{multirow}
\usepackage{ulem}
\usepackage{graphicx,color,dcolumn,booktabs,bm}
\usepackage{epsfig,dsfont,amssymb,amsmath,amsfonts,amsbsy,mathrsfs}

\graphicspath{{Figures/}} %

\usepackage{hyperref}
\hypersetup{colorlinks,citecolor=blue,anchorcolor=red,menucolor=red, linkcolor=red,filecolor=red,runcolor=red,urlcolor=blue,frenchlinks=red}

\makeatletter
\@addtoreset{equation}{section}
\makeatother

\allowdisplaybreaks

\begin{document}

\title{Going in quest of potential tetraquark interpretations for the newly observed $T_{\psi\psi}$ states in light of the diquark-antidiquark scenarios}
\author{Wen-Chao Dong$^{1,2}$}
\email{wen-chao$_$dong@qq.com}
\author{Zhi-Gang Wang$^{1}$\footnote{Corresponding author}}
\email{zgwang@aliyun.com}
\affiliation{$^1$Department of Physics, North China Electric Power University, Baoding 071003, China\\
$^2$School of Nuclear Science and Engineering, North China Electric Power University, Beijing 102206, China}

\begin{abstract}
Stimulated by the recent experimental progress on the $T_{\psi\psi}$ states, the fully charmed tetraquark spectroscopy is systemically investigated by dint of the Godfrey-Isgur relativized diquark model, the modified Godfrey-Isgur relativized diquark model with the color screening effects, and the nonrelativistic diquark model. The theoretical results of the diquark-antidiquark scenarios propose to interpret the $T_{\psi\psi}(6200)$, $T_{\psi\psi}(6600)$, $T_{\psi\psi}(6900)$, and $T_{\psi\psi}(7300)$ structures as the candidates of the $1S$-wave, $1P/2S$-wave, $1D/2P$-wave, and $2D/3P/4S$-wave fully charmed tetraquark states, respectively. On account of the deficiency of sufficient experimental information, e.g., the parities of the newly observed $T_{\psi\psi}$ states, there are uncertainties about the assignments of the $T_{\psi\psi}(6600)$, $T_{\psi\psi}(6900)$, and $T_{\psi\psi}(7300)$ states. It is demonstrated that the further experimental survey on the $cc\bar c\bar c$ states, implemented by the LHCb, ATLAS, CMS, and other collaborations, ought to be continued in the future.
\end{abstract}

\maketitle

\section{Introduction}\label{sec1}

Heavy flavored exotic hadrons discovered by experiments are sharply growing in recent two decades \cite{Workman:2022ynf,Olsen:2017bmm}. They comprise not only multitudinous hidden heavy flavored exotics, but also ample open heavy flavored exotics. For instance, the heavy quarkonium-like $T_\psi/T_\Upsilon$ states, the hidden charmed pentaquark $P_\psi/P_{\psi s}$ states, the open charmed tetraquark $T_{cs}/T_{cc}$ states, and so forth. What's more, the theoretical interpretations on the internal structure of heavy flavored exotics are also multifarious \cite{Hosaka:2016pey,Chen:2016qju,Ali:2017jda,Lebed:2016hpi,Esposito:2016noz,Guo:2017jvc,Karliner:2017qhf,Liu:2019zoy,Brambilla:2019esw,Chen:2022asf}, including compact multiquarks, loose hadronic molecules, hadro-quarkonia, hybrid hadrons, kinematical effects, etc. Nonetheless, the elucidation on the configurations of a portion of exotic states is in a dilemma due to the deficiency of experimental and theoretical smoking gun. Concretely, the conventional hadronic assignments for certain heavy flavored exotics cannot be ruled out, e.g., the $\chi_{c1}(3872)$ state. As a neutral state, it is capable of being regarded as the charmonium, the charmonium-like exotica, the mixing among them, or others. It demonstrates that the endeavor on shedding light on the nature of heavy flavored exotic hadrons is still herculean, despite progress.

Members of the $T_{\psi\psi}$ family are overtly exotic states beyond the conventional quark model, composed of four charm quarks ($cc\bar c\bar c$). The first observation on them was announced by the LHCb collaboration in 2020 \cite{LHCb:2020bwg}, with a narrow structure around 6.9 GeV in the $J/\psi$-pair (di-$J/\psi$) invariant mass spectrum, i.e., the $T_{\psi\psi}(6900)$ state. The corresponding detection was implemented by utilizing proton-proton collision at centre-of-mass energies of $\sqrt{s}=7$, 8, and 13 TeV, obtaining two sorts of possible alternatives for the mass and natural width of the $T_{\psi\psi}(6900)$ structure, enumerated in Table \ref{exp}. Whereafter, the presence of the $T_{\psi\psi}(6900)$ state was separately confirmed by the ATLAS and CMS collaborations in 2022 \cite{ATLAS:2022hhx,CMS:2022yhl}. Furthermore, a few else members of the $T_{\psi\psi}$ family, including the $T_{\psi\psi}(6200)$, $T_{\psi\psi}(6600)$, and $T_{\psi\psi}(7300)$ states, were not only recorded by the ATLAS collaboration in an excess of the di-$J/\psi$ and $J/\psi$+$\psi$(2S) events in the four-muon final state \cite{ATLAS:2022hhx}, but also observed by the CMS collaboration in the di-$J/\psi$ invariant mass spectrum produced by proton-proton collision data at $\sqrt{s}=13$ TeV \cite{CMS:2022yhl}. A series of discoveries of the $T_{\psi\psi}$ family members offered tremendous opportunities for the studies on the spectroscopic properties of heavy flavored exotic hadrons.

\renewcommand\tabcolsep{0.92cm}
\renewcommand{\arraystretch}{1.5}
\begin{table*}[!htbp]
\caption{A summary of experimental information of the $T_{\psi\psi}$ states which have been detected by the LHCb \cite{LHCb:2020bwg}, ATLAS \cite{ATLAS:2022hhx}, and CMS \cite{CMS:2022yhl} collaborations (in unit of MeV).}\label{exp}
\begin{tabular}{ccccc}
\toprule[1.0pt]\toprule[1.0pt]
Notation & Mass & Decay width & Channel & Experiment \\
\midrule[1.0pt]
$T_{\psi\psi}(6200)_{\rm A}$ & $6220\pm50^{+40}_{-50}$ & $310\pm120^{+70}_{-80}$ & $J/\psi$+$J/\psi$ & ATLAS \cite{ATLAS:2022hhx} \\
\midrule[0.5pt]
$T_{\psi\psi}(6600)_{\rm C}$ & $6552\pm10\pm12$ & $124\pm29\pm34$ & $J/\psi$+$J/\psi$ & CMS Fit I \cite{CMS:2022yhl} \\
$T_{\psi\psi}(6600)_{\rm A}$ & $6620\pm30^{+20}_{-10}$ & $310\pm90^{+60}_{-110}$ & $J/\psi$+$J/\psi$ & ATLAS \cite{ATLAS:2022hhx} \\
$T_{\psi\psi}(6600)_{\rm C'}$ & $6736\pm38$ & $439\pm65$ & $J/\psi$+$J/\psi$ & CMS Fit II \cite{CMS:2022yhl} \\
$T_{\psi\psi}(6600)_{\rm L'}$ & $6741\pm6$ & $288\pm16$ & $J/\psi$+$J/\psi$ & LHCb Fit II \cite{LHCb:2020bwg} \\
\midrule[0.5pt]
$T_{\psi\psi}(6900)_{\rm A'}$ & $6780\pm360^{+350}_{-540}$ & $390\pm110^{+110}_{-70}$ & $J/\psi$+$\psi$(2S) & ATLAS \cite{ATLAS:2022hhx} \\
$T_{\psi\psi}(6900)_{\rm A}$ & $6870\pm30^{+60}_{-10}$ & $120\pm40^{+30}_{-10}$ & $J/\psi$+$J/\psi$ & ATLAS \cite{ATLAS:2022hhx} \\
$T_{\psi\psi}(6900)_{\rm L'}$ & $6886\pm11\pm11$ & $168\pm33\pm69$ & $J/\psi$+$J/\psi$ & LHCb Fit II \cite{LHCb:2020bwg} \\
$T_{\psi\psi}(6900)_{\rm L}$ & $6905\pm11\pm7$ & $80\pm19\pm33$ & $J/\psi$+$J/\psi$ & LHCb Fit I \cite{LHCb:2020bwg} \\
$T_{\psi\psi}(6900)_{\rm C'}$ & $6918\pm10$ & $187\pm40$ & $J/\psi$+$J/\psi$ & CMS Fit II \cite{CMS:2022yhl} \\
$T_{\psi\psi}(6900)_{\rm C}$ & $6927\pm9\pm5$ & $122\pm22\pm19$ & $J/\psi$+$J/\psi$ & CMS Fit I \cite{CMS:2022yhl} \\
\midrule[0.5pt]
$T_{\psi\psi}(7300)_{\rm A'}$ & $7220\pm30^{+20}_{-30}$ & $100^{+130+60}_{-70-50}$ & $J/\psi$+$\psi$(2S) & ATLAS \cite{ATLAS:2022hhx} \\
$T_{\psi\psi}(7300)_{\rm C}$ & $7287\pm19\pm5$ & $95\pm46\pm20$ & $J/\psi$+$J/\psi$ & CMS \cite{CMS:2022yhl} \\
\bottomrule[1.0pt]\bottomrule[1.0pt]
\end{tabular}
\end{table*}

From the viewpoints of theoretical approaches, the existence of multiquark states is feasible, having been come up with by the pioneering works of the quark model \cite{Gell-Mann:1964ewy,Zweig:1964ruk}. In terms of the $T_{\psi\psi}$ family, the earliest theoretical inquiry was conducted by Ref. \cite{Iwasaki:1975pv} in 1975, predicting a sharp $cc\bar c\bar c$ resonance whose mass lay around 6.2 GeV. Over the ensuing decades, mass spectra, decay properties, and production mechanisms of the $cc\bar c\bar c$ states were systematically investigated by Refs. \cite{Chao:1980dv,Ader:1981db,Heller:1985cb,Badalian:1985es,Lloyd:2003yc,Barnea:2006sd,Chiu:2005ey,Berezhnoy:2011xn,Heupel:2012ua,Wang:2017jtz,
Chen:2016jxd,Karliner:2016zzc,Wu:2016vtq,Anwar:2017toa,Debastiani:2017msn,Liu:2019zuc,Wang:2018poa,Esau:2019hqw,Wang:2019rdo,Becchi:2020uvq,Gordillo:2020sgc,
Wang:2020gmd,Albuquerque:2020hio,Jin:2020jfc,Zhao:2020nwy,Karliner:2020dta,Lu:2020cns,Faustov:2020qfm,Lundhammar:2020xvw,Giron:2020wpx,Bedolla:2019zwg,Yang:2020atz,
Wang:2020ols,Richard:2020hdw,Maiani:2020pur,Chen:2020xwe,Chao:2020dml,Ma:2020kwb,Feng:2020riv,liu:2020eha,Zhang:2020xtb,Sonnenschein:2020nwn,Feng:2020qee,
Wang:2021mma,Yang:2021hrb,Wang:2020tpt,Goncalves:2021ytq,Zhu:2020xni,Wang:2021kfv,Liu:2021rtn,Liang:2021fzr,Huang:2021vtb,Guo:2020pvt,Pal:2021gkr,Li:2021ygk,
Mutuk:2021hmi,Maciula:2020wri,Wang:2020dlo,Yang:2020wkh,Cao:2020gul,Zhao:2020zjh,Weng:2020jao,Ke:2021iyh,Chen:2021crg,Deng:2020iqw,Nefediev:2021pww,Dong:2020nwy,
Dong:2021lkh,Wang:2020wrp,Faustov:2021hjs,Kuang:2022vdy,Wang:2022xja,Majarshin:2021hex,Chen:2022sbf,Santowsky:2021bhy,Wang:2022jmb,Zhuang:2021pci,Asadi:2021ids,
Gong:2020bmg,Wu:2022qwd,Gong:2022hgd,Tiwari:2021tmz,Wang:2022yes,Niu:2022cug,Faustov:2022mvs,Zhou:2022xpd,Zhang:2022qtp,An:2022qpt,Biloshytskyi:2022dmo,Pal:2023yid,
Kuang:2023vac,Yu:2022lak}, in virtue of miscellaneous phenomenological recipes, containing the quark bag model \cite{Chao:1980dv,Ader:1981db,Heller:1985cb}, the quark potential model \cite{Badalian:1985es,Lloyd:2003yc,Barnea:2006sd,Berezhnoy:2011xn,Karliner:2016zzc,Wu:2016vtq,Anwar:2017toa,Debastiani:2017msn,Liu:2019zuc,
Esau:2019hqw,Wang:2019rdo,Gordillo:2020sgc,Jin:2020jfc,Zhao:2020nwy,Karliner:2020dta,Lu:2020cns,Faustov:2020qfm,Lundhammar:2020xvw,Bedolla:2019zwg,Yang:2020atz,
Richard:2020hdw,liu:2020eha,Yang:2021hrb,Wang:2021kfv,Liu:2021rtn,Pal:2021gkr,Mutuk:2021hmi,Zhao:2020zjh,Weng:2020jao,Chen:2021crg,
Deng:2020iqw,Nefediev:2021pww,Faustov:2021hjs,Zhuang:2021pci,Asadi:2021ids,Tiwari:2021tmz,Wang:2022yes,Faustov:2022mvs,Zhang:2022qtp,An:2022qpt,Pal:2023yid,Yu:2022lak}, the Bethe-Salpeter (BS) equation \cite{Heupel:2012ua,Zhu:2020xni,Li:2021ygk,Ke:2021iyh,Santowsky:2021bhy}, the lattice QCD \cite{Chiu:2005ey}, the QCD sum rules \cite{Wang:2017jtz,Chen:2016jxd,Wang:2018poa,Albuquerque:2020hio,Wang:2020ols,Chen:2020xwe,Zhang:2020xtb,Wang:2021mma,Wang:2020dlo,Yang:2020wkh,Wang:2022xja,
Chen:2022sbf,Wu:2022qwd}, the adiabatic (Born-Oppenheimer) approximation \cite{Giron:2020wpx}, the nonrelativistic QCD (NRQCD) effective field theory \cite{Ma:2020kwb,Feng:2020riv,Feng:2020qee,Huang:2021vtb}, the holography inspired stringy hadron (HISH) model \cite{Sonnenschein:2020nwn}, the dynamical rescattering mechanism \cite{Wang:2020tpt,Wang:2020wrp,Wang:2022jmb}, the coupled-channel final state interaction (FSI) \cite{Liang:2021fzr,Guo:2020pvt,Cao:2020gul,Nefediev:2021pww,Dong:2020nwy,Dong:2021lkh,Zhuang:2021pci,Gong:2020bmg,Gong:2022hgd,Zhou:2022xpd,Kuang:2023vac}, the basis light front quantization (BLFQ) \cite{Kuang:2022vdy}, the bosonic algebraic approach \cite{Majarshin:2021hex}, etc. Since the $T_{\psi\psi}(6900)$ state was discovered by the LHCb collaboration \cite{LHCb:2020bwg}, numerous theoretical arguments on its inner composition were proposed, including the (diquark-antidiquark) tetraquark configuration \cite{Becchi:2020uvq,Gordillo:2020sgc,Albuquerque:2020hio,Jin:2020jfc,Zhao:2020nwy,Karliner:2020dta,Lu:2020cns,Faustov:2020qfm,Giron:2020wpx,Bedolla:2019zwg, Wang:2020ols,Richard:2020hdw,Maiani:2020pur,Chen:2020xwe,Chao:2020dml,liu:2020eha,Zhang:2020xtb,Sonnenschein:2020nwn,Zhu:2020xni,Wang:2021kfv,Liu:2021rtn,
Pal:2021gkr,Li:2021ygk,Mutuk:2021hmi,Wang:2020dlo,Zhao:2020zjh,Ke:2021iyh,Deng:2020iqw,Faustov:2021hjs,Wang:2022xja,Chen:2022sbf,Asadi:2021ids,
Tiwari:2021tmz,Faustov:2022mvs,Zhou:2022xpd,Zhang:2022qtp,An:2022qpt,Kuang:2023vac,Yu:2022lak}, the (meson-meson) molecule configuration \cite{Albuquerque:2020hio,Yang:2021hrb,Santowsky:2021bhy,Asadi:2021ids}, the (mixing among diquark-antidiquark and meson-meson) tetraquark-molecule mixing configuration \cite{Yang:2020atz,Chen:2021crg,Santowsky:2021bhy,Wang:2022yes}, the (hidden color-octet) molecule-like configuration \cite{Wang:2021mma,Yang:2020wkh}, the (coupled-channel) dynamically generated resonance pole structure \cite{Liang:2021fzr,Guo:2020pvt,Gong:2020bmg,Gong:2022hgd}, the (cusp) kinematical effect \cite{Zhuang:2021pci}, the light Higgs-like boson \cite{Zhu:2020snb}, the gluonic tetracharm configuration \cite{Wan:2020fsk}, and so on. Moreover, a vague structure around 7.2 GeV appeared in the di-$J/\psi$ invariant mass spectrum from the observation of the LHCb collaboration \cite{LHCb:2020bwg} could be construed as the $\bar\Xi_{cc}\Xi_{cc}$ molecular state with $J^{PC}=0^{-+}$ \cite{Wang:2021wjd} or the mixing among the (diquark-antidiquark) tetraquark state and the $\bar\Xi_{cc}\Xi_{cc}$ molecular state with $J^{PC}=0^{-+}$ \cite{Liu:2020tqy}. On account of the paucity of adequate experimental evidence, the nature of the $T_{\psi\psi}(6200)$, $T_{\psi\psi}(6600)$, $T_{\psi\psi}(6900)$, and $T_{\psi\psi}(7300)$ states has all been intangible so far.

Howbeit, the (diquark-antidiquark) tetraquark configurations are deemed as the quite probable interpretations for the $T_{\psi\psi}$ structures which have been reported by experiments, owing to the absence of the light quark degree of freedom, which plays an imperative role in the formation of (meson-meson) molecular states \cite{Maiani:2020pur,Chao:2020dml}. Consequently, a few diquark-antidiquark scenarios are employed to delve into the spectroscopic properties of the $cc\bar c\bar c$ states in this work, covering the Godfrey-Isgur (GI) relativized diquark model, the modified Godfrey-Isgur (MGI) relativized diquark model (incorporating the color screening effects), and the nonrelativistic (NR) diquark model. The precis of this treatise is recapitulated as follows. Initially, the status quo of the experimental and theoretical investigations on heavy flavored exotic hadrons is laconically retrospected in Section \ref{sec1}. Subsequently, three sorts of diquark-antidiquark scenarios are expounded in Section \ref{sec2}. Afterwards, the mass spectrum of the low-lying excited $T_{\psi\psi}$ family is procured in Section \ref{sec3}. Next, both the potential interpretations of the newly observed $T_{\psi\psi}$ structures and the spectroscopic discrepancies between this work and other theoretical approaches are articulated in Section \ref{sec4}. Lastly, a compendious summary of this work is presented in Section \ref{sec5}.

\section{Formalism}\label{sec2}

This section mainly delineates the Godfrey-Isgur (GI) relativized diquark model, the modified Godfrey-Isgur (MGI) relativized diquark model with the color screening effects, the nonrelativistic (NR) diquark model, and their applications on the inquiries on the spectroscopic properties of quadruply charmed (diquark-antidiquark) tetraquarks.

\subsection{Godfrey-Isgur (GI) relativized diquark model}\label{subsec21}

The Godfrey-Isgur relativized quark model (GI model) was proposed by S. Godfrey and N. Isgur in 1985 \cite{Godfrey:1985xj}, studying the spectroscopic properties of all sorts of mesons with a set of universal parameters of the one-gluon-exchange-plus-linear-confinement potential incited by QCD. The pivotal feature of the GI model is, as Ref. \cite{Godfrey:1985xj} narrates, ``\textit{all mesons---from the pion to the upsilon---can be described in a unified framework}''. Thus far, the GI model has been successfully employed to probe the mass spectra of light mesons \cite{Godfrey:1985xj}, heavy-light mesons \cite{Godfrey:1985xj,Godfrey:2016nwn,Godfrey:2015dva}, fully heavy mesons \cite{Godfrey:1985xj,Godfrey:2004ya,Barnes:2005pb,Godfrey:2015dia}, light baryons \cite{Capstick:1986ter}, singly heavy baryons \cite{Capstick:1986ter,Lu:2016ctt}, doubly heavy baryons \cite{Lu:2017meb}, fully heavy baryons \cite{Bedolla:2019zwg}, light tetraquarks \cite{Lu:2019ira}, heavy tetraquarks \cite{Lu:2016zhe,Lu:2016cwr}, and diquarks \cite{Anwar:2017toa,Bedolla:2019zwg,Lu:2016ctt,Lu:2019ira,Lu:2016zhe,Lu:2016cwr,Ferretti:2019zyh}. The specific details and explicit expressions of the GI model may be found in Ref. \cite{Godfrey:1985xj}. As far as meson is concerned, the Hamiltonian of the GI model is
\begin{eqnarray}
H_{\rm GI}&=&H_{\rm GI}^0+V_{\rm GI}^{\rm conf}+V_{\rm GI}^{\rm cont}+V_{\rm GI}^{\rm ten}+V_{\rm GI}^{\rm so},\label{eq21}
\end{eqnarray}
where $H_{\rm GI}^0$ is the relativistic energy of total (anti)quarks. $V_{\rm GI}^{\rm conf}$, $V_{\rm GI}^{\rm cont}$, $V_{\rm GI}^{\rm ten}$, and $V_{\rm GI}^{\rm so}$ denote the effective confinement, contact, tensor, and spin-orbit potentials between the quark 1 and antiquark 2 in the meson, respectively. Among them, $V_{\rm GI}^{\rm ten}$ and $V_{\rm GI}^{\rm so}$ are capable of being decomposed into the diagonal terms ($V_{\rm GI}^{\rm (ten)}$ and $V_{\rm GI}^{\rm (so)}$) and off-diagonal terms ($V_{\rm GI}^{\rm [ten]}$ and $V_{\rm GI}^{\rm [so]}$). In this work, both of off-diagonal terms ($V_{\rm GI}^{\rm [ten]}$ and $V_{\rm GI}^{\rm [so]}$) are omitted for convenience. As a matter of fact, the spin-orbit off-diagonal term $V_{\rm GI}^{\rm [so]}$ is going to vanish automatically when the mass of quark 1 is equivalent to the mass of antiquark 2 \cite{Godfrey:1985xj}. The forms of $H_{\rm GI}^0$, $V_{\rm GI}^{\rm conf}$, $V_{\rm GI}^{\rm cont}$, $V_{\rm GI}^{\rm (ten)}$, and $V_{\rm GI}^{\rm (so)}$ are
\begin{eqnarray}
H_{\rm GI}^0&=&\sum_{i=1}^2E_{i}(p),\label{eq22}\\
V_{\rm GI}^{\rm conf}&=&\tilde G_{12}^{\rm Coul}(p,r)+\tilde S_{12}(r),\label{eq23}\\
V_{\rm GI}^{\rm cont}&=&\frac{2}{3m_1m_2r^2}\frac{\partial}{\partial r}\left[r^2\frac{\partial\tilde G_{12}^{\rm cont}(p,r)}{\partial r}\right]\bm S_1\cdot\bm S_2,\label{eq24}\\
V_{\rm GI}^{\rm (ten)}&=&\frac{1}{m_1m_2}\left(\frac{1}{r}-\frac{\partial}{\partial r}\right)\frac{\partial\tilde G_{12}^{\rm ten}(p,r)}{\partial r}\mathbb T_{\rm diag},\label{eq25}\\
V_{\rm GI}^{\rm (so)}&=&\frac{1}{r}\left[\frac{1}{4m_1^2}\frac{\partial\tilde G_{11}^{\rm so(v)}(p,r)}{\partial r}+\frac{1}{4m_2^2}\frac{\partial\tilde G_{22}^{\rm so(v)}(p,r)}{\partial r}\right.\nonumber\\
&&-\frac{1}{4m_1^2}\frac{\partial\tilde S_{11}^{\rm so(s)}(p,r)}{\partial r}-\frac{1}{4m_2^2}\frac{\partial\tilde S_{22}^{\rm so(s)}(p,r)}{\partial r}\nonumber\\
&&\left.+\frac{1}{m_1m_2}\frac{\partial\tilde G_{12}^{\rm so(v)}(p,r)}{\partial r}\right]\bm L\cdot\bm S,\label{eq26}
\end{eqnarray}
with
\begin{eqnarray*}
&&E_{i}(p)=\left(p^2+m_i^2\right)^\frac{1}{2},\\
&&\tilde G_{ij}^{\rm Coul}(p,r)\nonumber\\
&=&\left[1+\frac{p^2}{E_i(p)E_j(p)}\right]^\frac{1}{2}\tilde G_{ij}(r)\left[1+\frac{p^2}{E_i(p)E_j(p)}\right]^\frac{1}{2},\\
&&\tilde G_{ij}^{\rm cont/ten/so(v)}(p,r)\nonumber\\
&=&\left[\frac{m_im_j}{E_i(p)E_j(p)}\right]^{\frac{1}{2}+\epsilon_{\rm cont/ten/so(v)}}\tilde G_{ij}(r)\left[\frac{m_im_j}{E_i(p)E_j(p)}\right]^{\frac{1}{2}+\epsilon_{\rm cont/ten/so(v)}},\\
&&\tilde S_{ij}^{\rm so(s)}(p,r)\nonumber\\
&=&\left[\frac{m_im_j}{E_i(p)E_j(p)}\right]^{\frac{1}{2}+\epsilon_{\rm so(s)}}\tilde S_{ij}(r)\left[\frac{m_im_j}{E_i(p)E_j(p)}\right]^{\frac{1}{2}+\epsilon_{\rm so(s)}},\\
&&\mathbb T=\frac{\mathbb S_{12}}{12}=\frac{\left(\bm S_1\cdot\bm r\right)\left(\bm S_2\cdot\bm r\right)}{r^2}-\frac{1}{3}\bm S_1\cdot\bm S_2.
\end{eqnarray*}
Here, $\mathbb T$ denotes the operator of the tensor coupling interaction, whose diagonal matrix elements are able to be evaluated by dint of the identity from Landau and Lifshitz \cite{Landau:1991wop,Ali:2017wsf} or the Wigner-Eckart theorem \cite{Varshalovich:1988krb}. $E_{i}$ and $m_{i}$ are the relativistic energy and mass of the (anti)quark $i$, respectively. In consideration of the various types of potentials' dependence on the center-of-mass momentum of the interacting (anti)quarks, the momentum-dependent factors are introduced into the smeared Coulomb and linear potentials ($\tilde G_{ij}$ and $\tilde S_{ij}$), to acquire the corresponding momentum-dependent Coulomb, contact, tensor, vector spin-orbit, and scalar spin-orbit potentials ($\tilde G_{ij}^{\rm Coul}$, $\tilde G_{ij}^{\rm cont}$, $\tilde G_{ij}^{\rm ten}$, $\tilde G_{ij}^{\rm so(v)}$, and $\tilde S_{ij}^{\rm so(s)}$), with the universal parameters ($\epsilon_{\rm cont}$, $\epsilon_{\rm ten}$, $\epsilon_{\rm so(v)}$, and $\epsilon_{\rm so(s)}$) \cite{Godfrey:1985xj}. The smearing prescription of the certain potential $f(r)$ is defined as \cite{Godfrey:1985xj}
\begin{eqnarray}
\tilde f_{ij}(r)&\equiv&\int d^3r'\rho_{ij}({\bf r}-{\bf r}')f(r'),\label{eq27}
\end{eqnarray}
with
\begin{gather}
\rho_{ij}({\bf r}-{\bf r}')=\frac{\sigma_{ij}^3}{\pi^\frac{3}{2}}e^{-\sigma_{ij}^2({\bf r}-{\bf r}')^2},\label{eq28}\\
\sigma_{ij}=\sqrt{\sigma_0^2\left[\frac{1}{2}+\frac{1}{2}\left(\frac{4m_im_j}{(m_i+m_j)^2}\right)^4\right]+s^2\left(\frac{2m_im_j}{m_i+m_j}\right)^2}.\label{eq29}
\end{gather}
Here, $\rho_{ij}({\bf r}-{\bf r}')$ is the smearing function, with the universal parameters ($\sigma_0$ and $s$) \cite{Godfrey:1985xj}. By taking advantage of Eq. (\ref{eq27}), the short-range $\gamma^\mu\otimes\gamma_\mu$ Coulomb potential $G(r)$ and long-range $1\otimes1$ linear potential $S(r)$ are capable of being smeared into the smeared Coulomb potential $\tilde G_{ij}(r)$ and smeared linear potential $\tilde S_{ij}(r)$, respectively. The forms of $G(r)$ and $S(r)$ are
\begin{eqnarray}
G(r)&=&\frac{\alpha_s(r)}{4r}\bm\lambda_1\cdot\bm\lambda_2,\label{eq210}\\
S(r)&=&-\frac{3}{16}(br+c)\bm\lambda_1\cdot\bm\lambda_2,\label{eq211}
\end{eqnarray}
with
\begin{eqnarray}
\alpha_s(r)&=&\sum_{k=1}^3\alpha_k{\rm erf}(\gamma_kr),\label{eq212}\\
{\rm erf}(x)&=&\frac{2}{\pi^\frac{1}{2}}\int_0^xe^{-t^2}dt.\label{eq213}
\end{eqnarray}
Here, $\bm\lambda_i$ denotes the Gell-Mann matrices of the color SU(3) group, which act on the (anti)quark $i$. The matrix elements of $\bm\lambda_i\cdot\bm\lambda_j$ are able to be evaluated in virtue of the eigenvalue of the quadratic Casimir operator \cite{Vijande:2009ac}. ${\rm erf}(x)$ and aforementioned constants ($\alpha_{k=1,2,3}$, $\gamma_{k=1,2,3}$, $b$, and $c$) are the error function and GI model universal parameters \cite{Godfrey:1985xj}, respectively.

As the hypothetical substructure of the baryon and multiquark state, composed of two quarks, the diquark plays a crucial role in perceiving the various properties of hadrons \cite{Anselmino:1992vg,Jaffe:2004ph,Barabanov:2020jvn}, including but not limited to the spectroscopy, structure, form factors, magnetic moments, production, and decay properties of hadrons. The representation of the color SU(3) group of the diquark is antitriplet or sextet inasmuch as the quark possesses the color triplet representation, i.e.,
\begin{eqnarray}
{\bf 3}_q\otimes{\bf 3}_q&=&\bar{\bf 3}_{qq}\oplus{\bf 6}_{qq},\label{eq214}\\
\bar{\bf 3}_{\bar q}\otimes\bar{\bf 3}_{\bar q}&=&{\bf 3}_{\bar q\bar q}\oplus\bar{\bf 6}_{\bar q\bar q}.\label{eq215}
\end{eqnarray}
What's remarkable is that the matrix element $\langle\bm\lambda_1\cdot\bm\lambda_2\rangle$ of the color (anti)sextet (anti)diquark is positive $4/3$, which means that the quark-quark (antiquark-antiquark) interaction in the color (anti)sextet (anti)diquark is repulsive, engendering that the color (anti)sextet (anti)diquark is incapable of being formed in the diquark model \cite{Bedolla:2019zwg,Ke:2021iyh,Lu:2019ira,Lu:2016zhe,Lu:2016cwr,Ferretti:2019zyh}. Hence, the diquark-antidiquark scenarios employed in this work only adopt the color antitriplet diquark (triplet antidiquark) as the practically effective (anti)diquark to investigate spectroscopy of fully charmed tetraquarks. Manifestly, the color antitriplet diquark (triplet antidiquark) is able to be approximately deemed as the antiquark (quark), since it possesses the color representation in common with the antiquark (quark). The GI relativized diquark model is carried out in two steps. First of all, the mass of the ground state doubly charmed diquark is garnered. Next up, the spectroscopic properties of quadruply charmed tetraquarks are surveyed via looking upon the diquark (antidiquark) as the antiquark (quark). All the corresponding results are laid out in Section \ref{sec3}.

\subsection{Modified Godfrey-Isgur (MGI) relativized diquark model with the color screening effects}\label{subsec22}

Admittedly, the linear potential $S(r)$ is quite successful for depicting the long-range interaction between (anti)quarks \cite{Godfrey:1985xj}. However, the color flux tube between (anti)quarks may break up at large distances due to the vacuum polarization effects of dynamical fermions, which gives rise to the color screening effects \cite{Born:1989iv}. The color screened nonrelativistic quark potential models have been successfully employed to study the mass spectrum of heavy quarkonium \cite{Li:2009zu}, by means of superseding the linear potential $S(r)$ with the screened linear potential $S^{\rm scr}(r)$, which is
\begin{eqnarray}
S^{\rm scr}(r)&=&-\frac{3}{16}\left(b\frac{1-e^{-\mu r}}{\mu}+c\right)\bm\lambda_1\cdot\bm\lambda_2.\label{eq216}
\end{eqnarray}
Here, $\mu$ is the color screening factor. It is straightforward to be aware of
\begin{eqnarray}
S^{\rm scr}(r)\rightarrow\left\{
\begin{aligned}
S(r)=&-\frac{3}{16}(br+c)\bm\lambda_1\cdot\bm\lambda_2,&&r\rightarrow0,\\
c_\mu=&-\frac{3}{16}\left(\frac{b}{\mu}+c\right)\bm\lambda_1\cdot\bm\lambda_2,&&r\rightarrow\infty,\label{eq217}
\end{aligned}
\right.
\end{eqnarray}
which indicates that the screened linear potential $S^{\rm scr}(r)$ inclines to the linear potential $S(r)$ when the interaction range $r$ verges on zero, and inclines to the constant $c_\mu$ when the interaction range $r$ verges on infinity.

By incorporating the color screening effects, the modified Godfrey-Isgur relativized quark model (MGI model) is attained \cite{Song:2015nia}, having been successfully employed to investigate the spectroscopic properties of light mesons \cite{Pang:2017dlw}, heavy-light mesons \cite{Song:2015nia,Song:2015fha}, fully heavy mesons \cite{Wang:2018rjg}, light tetraquarks \cite{Lu:2019ira}, heavy tetraquarks \cite{Lu:2016cwr}, and diquarks \cite{Lu:2019ira,Lu:2016cwr}. In the MGI model, the smeared screened linear potential $\tilde S_{ij}^{\rm scr}(r)$ is procured in virtue of interpolating the screened linear potential $S^{\rm scr}(r)$ into Eq. (\ref{eq27}). Besides, the other particulars of the MGI model are in concordance with the GI model. The specific details and explicit expressions of the MGI model may be found in Ref. \cite{Song:2015nia}. Whereafter, the MGI relativized diquark model is performed in two aforementioned steps, which garners the masses of the doubly charmed diquark and fully charmed tetraquarks displayed in Section \ref{sec3}.

\subsection{Nonrelativistic (NR) diquark model}\label{subsec23}

As one of the potent phenomenological approaches, the Cornell potential model is an eminent exemplar of the nonrelativistic quark potential models, limning the mass spectrum of quarkonium successfully \cite{Eichten:1979ms}. It is composed of the Cornell Coulomb potential $G^{\rm Corn}(r)$ and Cornell linear potential $S^{\rm Corn}(r)$, i.e.,
\begin{eqnarray}
G^{\rm Corn}(r)&=&\frac{\alpha_c}{4r}\bm\lambda_1\cdot\bm\lambda_2,\label{eq218}\\
S^{\rm Corn}(r)&=&-\frac{3}{16}\beta r\bm\lambda_1\cdot\bm\lambda_2.\label{eq219}
\end{eqnarray}

The nonrelativistic quark model (NR model) employed in this work stems from Ref. \cite{Barnes:2005pb}, having successfully inquired into the spectroscopy of the charmonium system. Its Hamiltonian is
\begin{eqnarray}
H_{\rm NR}&=&H_{\rm NR}^0+V_{\rm NR}^{\rm conf}+V_{\rm NR}^{\rm cont}+V_{\rm NR}^{\rm ten}+V_{\rm NR}^{\rm so},\label{eq220}
\end{eqnarray}
where the tensor and spin-orbit off-diagonal terms ($V_{\rm NR}^{\rm [ten]}$ and $V_{\rm NR}^{\rm [so]}$) are omitted as mentioned previously. Concretely, all the relevant terms of $H_{\rm NR}$ are
\begin{eqnarray}
H_{\rm NR}^0&=&\sum_{i=1}^2\mathcal E_{i}(p),\label{eq221}\\
V_{\rm NR}^{\rm conf}&=&G^{\rm Corn}(r)+S^{\rm Corn}(r),\label{eq222}\\
V_{\rm NR}^{\rm cont}&=&\frac{2}{3m_1m_2r^2}\frac{\rm d}{{\rm d}r}\left[r^2\frac{{\rm d}G^{\rm erf}(r)}{{\rm d}r}\right]\bm S_1\cdot\bm S_2,\label{eq223}\\
V_{\rm NR}^{\rm (ten)}&=&\frac{1}{m_1m_2}\left(\frac{1}{r}-\frac{\rm d}{{\rm d}r}\right)\frac{{\rm d}G^{\rm Corn}(r)}{{\rm d}r}\mathbb T_{\rm diag},\label{eq224}\\
V_{\rm NR}^{\rm (so)}&=&\frac{1}{r}\left[\left(\frac{1}{4m_1^2}+\frac{1}{4m_2^2}+\frac{1}{m_1m_2}\right)\frac{{\rm d}G^{\rm Corn}(r)}{{\rm d}r}\right.\nonumber\\
&&\left.-\left(\frac{1}{4m_1^2}+\frac{1}{4m_2^2}\right)\frac{{\rm d}S^{\rm Corn}(r)}{{\rm d}r}\right]\bm L\cdot\bm S,\label{eq225}
\end{eqnarray}
with
\begin{eqnarray}
\mathcal E_{i}(p)&=&m_i+\frac{p^2}{2m_i},\label{eq226}\\
G^{\rm erf}(r)&=&\frac{\mathfrak{\alpha}_e(r)}{4r}\bm\lambda_1\cdot\bm\lambda_2,\label{eq227}\\
\alpha_e(r)&=&\alpha_c{\rm erf}(\gamma_cr).\label{eq228}
\end{eqnarray}
Here, $\mathcal E_{i}$ and aforementioned constants ($\alpha_c$, $\gamma_c$, and $\beta$) are the nonrelativistic energy of the (anti)quark $i$ and NR model parameters determined by fitting the mass spectrum of the charmonium states \cite{Barnes:2005pb}, respectively. In the NR model, the form of $\alpha_e(r)$ in $G^{\rm erf}(r)$ employs a approximate form of $\alpha_s(r)$ in $G(r)$. Following Ref. \cite{Barnes:2005pb}, the tensor and spin-orbit terms ($V_{\rm NR}^{\rm (ten)}$ and $V_{\rm NR}^{\rm (so)}$) are deemed as mass shifts in virtue of leading-order perturbation theory. The specific details and explicit expressions of the NR model may be found in Ref. \cite{Barnes:2005pb}. Subsequently, the NR diquark model is implemented by adopting the (diquark-antidiquark) tetraquark configuration, whose results on the axial-vector doubly charmed diquark and fully charmed tetraquark system are enumerated in Section \ref{sec3}.

\section{Results}\label{sec3}

The particular results on the spectroscopic properties of the ground state axial-vector doubly charmed diquark and entire fully charmed tetraquark family are unveiled by three sorts of aforementioned diquark-antidiquark scenarios (GI relativized diquark model, MGI relativized diquark model, and NR diquark model) in this section.

\subsection{Parameters}\label{subsec31}

In the calculations on the mass spectra of doubly charmed diquark and fully charmed tetraquark states, fulfilled by the GI (MGI) relativized diquark model and NR diquark model, it's worth noting that the values of all the corresponding parameters are identical with the counterparts of Refs. \cite{Godfrey:1985xj,Barnes:2005pb}, as enumerated in Table \ref{para}.

\renewcommand\tabcolsep{0.15cm}
\renewcommand{\arraystretch}{1.5}
\begin{table}[!htbp]
\caption{Parameters of the GI (MGI) \cite{Godfrey:1985xj} and NR \cite{Barnes:2005pb} models.}\label{para}
\begin{tabular}{cccc|cc}
\toprule[1.0pt]\toprule[1.0pt]
\multicolumn{4}{c|}{GI (MGI) \cite{Godfrey:1985xj}} & \multicolumn{2}{c}{NR \cite{Barnes:2005pb}} \\
Parameter & Value & Parameter & Value & Parameter & Value \\
\midrule[1.0pt]
$m_c$ (GeV) & $1.628$ & $\cdots$ & $\cdots$ & $m_c$ (GeV) & $1.4794$ \\
$b$ (GeV$^2$) & $0.18$ & $c$ (GeV) & $-0.253$ & $\beta$ (GeV$^2$) & $0.1425$ \\
$\gamma_1$ (GeV) & $\sqrt{1/4}$ & $\alpha_1$ & $0.25$ & $\gamma_c$ (GeV) & $1.0946$ \\
$\gamma_2$ (GeV) & $\sqrt{10/4}$ & $\alpha_2$ & $0.15$ & $\alpha_c$ & $0.5461$ \\
$\gamma_3$ (GeV) & $\sqrt{1000/4}$ & $\alpha_3$ & $0.20$ & $\cdots$ & $\cdots$ \\
$\sigma_0$ (GeV) & $1.80$ & $s$ & $1.55$ & $\cdots$ & $\cdots$ \\
$\epsilon_{\rm cont}$ & $-0.168$ & $\epsilon_{\rm so(v)}$ & $-0.035$ & $\cdots$ & $\cdots$ \\
$\epsilon_{\rm ten}$ & $0.025$ & $\epsilon_{\rm so(s)}$ & $0.055$ & $\cdots$ & $\cdots$ \\
\bottomrule[1.0pt]\bottomrule[1.0pt]
\end{tabular}
\end{table}

\subsection{Doubly charmed diquark}\label{subsec32}

According to the Pauli exclusion principle, the total wave function of the (anti)diquark, made up of the color, flavor, spin, and spatial wave functions, is sure to be antisymmetric \cite{Amsler:2018zkm}. Ergo, in terms of the color antitriplet (triplet) ground state doubly charmed diquark (antidiquark), the spin wave function is symmetric, i.e., the spin quantum number has to be $1$. Thereby the color antitriplet (triplet) ground state doubly charmed diquark (antidiquark) must be an axial-vector (anti)diquark owing to the fact that the $S$-wave (anti)diquark possesses a positive internal parity \cite{Amsler:2018zkm}. Whereafter, the masses of the ground state axial-vector doubly charmed diquark, garnered by aforementioned diquark-antidiquark scenarios, are distinctly laid out in Table \ref{diquark}. In this work, the values of diquark masses, obtained by the GI, MGI ($\mu=50$), MGI ($\mu=100$), and NR models, are 3329, 3314, 3300, and 3152 MeV, respectively.

\renewcommand\tabcolsep{0.54cm}
\renewcommand{\arraystretch}{1.5}
\begin{table}[!htbp]
\caption{The ground state axial-vector doubly charmed diquark masses reaped by this work (in unit of MeV).}\label{diquark}
\begin{tabular}{cccc}
\toprule[1.0pt]\toprule[1.0pt]
Scenario & Mass & Scenario & Mass \\
\midrule[1.0pt]
GI & $3329$ & MGI ($\mu=50$) & $3314$ \\
NR & $3152$ & MGI ($\mu=100$) & $3300$ \\
\bottomrule[1.0pt]\bottomrule[1.0pt]
\end{tabular}
\end{table}

\subsection{Fully charmed tetraquark}\label{subsec33}

\renewcommand\tabcolsep{0.295cm}
\renewcommand{\arraystretch}{1.5}
\begin{table*}[!htbp]
\caption{The mass spectrum of the $1S$-, $1P$-, $2S$-, $1D$-, and $2P$-wave fully charmed tetraquark states reaped by this work (in unit of MeV).}\label{cccc1}
\begin{tabular}{cc|cc|cc|cc|cc}
\toprule[1.0pt]\toprule[1.0pt]
\multicolumn{2}{c|}{State} & \multicolumn{2}{c|}{GI} & \multicolumn{2}{c|}{MGI ($\mu=50$)} & \multicolumn{2}{c|}{MGI ($\mu=100$)} & \multicolumn{2}{c}{NR} \\
$T_{\psi\psi}(n^{2S+1}L_J)$ & $J^{PC}$ & Mass & Candidate & Mass & Candidate & Mass & Candidate & Mass & Candidate \\
\midrule[1.0pt]
$T_{\psi\psi0}^f(1^1S_0)$ & $0^{++}$ & $6053$ &  & $6020$ &  & $5989$ &  & $5944$ &  \\
$T_{\psi\psi1}^h(1^3S_1)$ & $1^{+-}$ & $6181$ &  & $6148$ &  & $6115$ &  & $6001$ &  \\
$T_{\psi\psi2}^f(1^5S_2)$ & $2^{++}$ & $6331$ &  & $6295$ &  & $6260$ & $T_{\psi\psi}(6200)_{\rm A}$ & $6105$ &  \\
$T_{\psi\psi0}^\eta(1^3P_0)$ & $0^{-+}$ & $6633$ & $T_{\psi\psi}(6600)_{\rm A}$ & $6588$ &  & $6545$ & $T_{\psi\psi}(6600)_{\rm C}$ & $6478$ &  \\
$T_{\psi\psi1}^\eta(1^3P_1)$ & $1^{-+}$ & $6697$ &  & $6650$ & $T_{\psi\psi}(6600)_{\rm A}$ & $6604$ & $T_{\psi\psi}(6600)_{\rm A}$ & $6584$ &  \\
$T_{\psi\psi1}^\omega(1^1P_1)$ & $1^{--}$ & $6698$ &  & $6651$ &  & $6605$ &  & $6584$ &  \\
$T_{\psi\psi1}^\omega(1^5P_1)$ & $1^{--}$ & $6633$ &  & $6588$ &  & $6544$ &  & $6495$ &  \\
$T_{\psi\psi2}^\eta(1^3P_2)$ & $2^{-+}$ & $6718$ & $T_{\psi\psi}(6600)_{\rm C'}$ & $6670$ &  & $6623$ & $T_{\psi\psi}(6600)_{\rm A}$ & $6618$ & $T_{\psi\psi}(6600)_{\rm A}$ \\
$T_{\psi\psi2}^\omega(1^5P_2)$ & $2^{--}$ & $6712$ &  & $6664$ &  & $6618$ &  & $6609$ &  \\
$T_{\psi\psi3}^\omega(1^5P_3)$ & $3^{--}$ & $6739$ &  & $6691$ &  & $6643$ &  & $6648$ &  \\
$T_{\psi\psi0}^f(2^1S_0)$ & $0^{++}$ & $6751$ & $T_{\psi\psi}(6600)_{\rm C'}$ & $6697$ &  & $6644$ & $T_{\psi\psi}(6600)_{\rm A}$ & $6667$ &  \\
$T_{\psi\psi1}^h(2^3S_1)$ & $1^{+-}$ & $6794$ &  & $6738$ &  & $6683$ &  & $6679$ &  \\
$T_{\psi\psi2}^f(2^5S_2)$ & $2^{++}$ & $6864$ & $T_{\psi\psi}(6900)_{\rm A}$ & $6803$ &  & $6744$ & $T_{\psi\psi}(6600)_{\rm C';L'}$ & $6703$ & $T_{\psi\psi}(6600)_{\rm C'}$ \\
$T_{\psi\psi0}^f(1^5D_0)$ & $0^{++}$ & $6953$ &  & $6891$ & $T_{\psi\psi}(6900)_{\rm A;L'}$ & $6831$ &  & $6826$ &  \\
$T_{\psi\psi1}^f(1^5D_1)$ & $1^{++}$ & $6963$ &  & $6900$ & $T_{\psi\psi}(6900)_{\rm A;L}$ & $6839$ &  & $6835$ &  \\
$T_{\psi\psi1}^h(1^3D_1)$ & $1^{+-}$ & $6970$ &  & $6907$ &  & $6846$ &  & $6841$ &  \\
$T_{\psi\psi2}^f(1^1D_2)$ & $2^{++}$ & $6985$ &  & $6922$ & $T_{\psi\psi}(6900)_{\rm C;C'}$ & $6860$ & $T_{\psi\psi}(6900)_{\rm A}$ & $6859$ & $T_{\psi\psi}(6900)_{\rm A}$ \\
$T_{\psi\psi2}^f(1^5D_2)$ & $2^{++}$ & $6978$ &  & $6915$ & $T_{\psi\psi}(6900)_{\rm L;C'}$ & $6853$ & $T_{\psi\psi}(6900)_{\rm A}$ & $6850$ & $T_{\psi\psi}(6900)_{\rm A}$ \\
$T_{\psi\psi2}^h(1^3D_2)$ & $2^{+-}$ & $6986$ &  & $6923$ &  & $6860$ &  & $6860$ &  \\
$T_{\psi\psi3}^f(1^5D_3)$ & $3^{++}$ & $6994$ &  & $6929$ & $T_{\psi\psi}(6900)_{\rm C}$ & $6867$ & $T_{\psi\psi}(6900)_{\rm A}$ & $6867$ & $T_{\psi\psi}(6900)_{\rm A}$ \\
$T_{\psi\psi3}^h(1^3D_3)$ & $3^{+-}$ & $6994$ &  & $6930$ &  & $6867$ &  & $6867$ &  \\
$T_{\psi\psi4}^f(1^5D_4)$ & $4^{++}$ & $7002$ &  & $6938$ &  & $6875$ & $T_{\psi\psi}(6900)_{\rm A;L'}$ & $6876$ & $T_{\psi\psi}(6900)_{\rm A;L'}$ \\
$T_{\psi\psi0}^\eta(2^3P_0)$ & $0^{-+}$ & $7050$ &  & $6975$ &  & $6902$ & $T_{\psi\psi}(6900)_{\rm L}$ & $6867$ & $T_{\psi\psi}(6900)_{\rm A}$ \\
$T_{\psi\psi1}^\eta(2^3P_1)$ & $1^{-+}$ & $7091$ &  & $7013$ &  & $6937$ &  & $6951$ &  \\
$T_{\psi\psi1}^\omega(2^1P_1)$ & $1^{--}$ & $7091$ &  & $7013$ &  & $6937$ &  & $6951$ &  \\
$T_{\psi\psi1}^\omega(2^5P_1)$ & $1^{--}$ & $7051$ &  & $6975$ &  & $6902$ &  & $6877$ &  \\
$T_{\psi\psi2}^\eta(2^3P_2)$ & $2^{-+}$ & $7106$ &  & $7027$ &  & $6949$ &  & $6977$ &  \\
$T_{\psi\psi2}^\omega(2^5P_2)$ & $2^{--}$ & $7102$ &  & $7023$ &  & $6946$ &  & $6970$ &  \\
$T_{\psi\psi3}^\omega(2^5P_3)$ & $3^{--}$ & $7122$ &  & $7041$ &  & $6963$ &  & $7002$ &  \\
\bottomrule[1.0pt]\bottomrule[1.0pt]
\end{tabular}
\end{table*}

\renewcommand\tabcolsep{0.345cm}
\renewcommand{\arraystretch}{1.5}
\begin{table*}[!htbp]
\caption{The mass spectrum of the $3S$-, $1F$-, $2D$-, $3P$-, and $4S$-wave fully charmed tetraquark states reaped by this work (in unit of MeV).}\label{cccc2}
\begin{tabular}{cc|cc|cc|cc|cc}
\toprule[1.0pt]\toprule[1.0pt]
\multicolumn{2}{c|}{State} & \multicolumn{2}{c|}{GI} & \multicolumn{2}{c|}{MGI ($\mu=50$)} & \multicolumn{2}{c|}{MGI ($\mu=100$)} & \multicolumn{2}{c}{NR} \\
$T_{\psi\psi}(n^{2S+1}L_J)$ & $J^{PC}$ & Mass & Candidate & Mass & Candidate & Mass & Candidate & Mass & Candidate \\
\midrule[1.0pt]
$T_{\psi\psi0}^f(3^1S_0)$ & $0^{++}$ & $7152$ &  & $7064$ &  & $6979$ &  & $7031$ &  \\
$T_{\psi\psi1}^h(3^3S_1)$ & $1^{+-}$ & $7180$ &  & $7089$ &  & $7001$ &  & $7038$ &  \\
$T_{\psi\psi2}^f(3^5S_2)$ & $2^{++}$ & $7230$ & $T_{\psi\psi}(7300)_{\rm A'}$ & $7134$ &  & $7040$ &  & $7054$ &  \\
$T_{\psi\psi1}^\omega(1^5F_1)$ & $1^{--}$ & $7197$ &  & $7115$ &  & $7035$ &  & $7046$ &  \\
$T_{\psi\psi2}^\eta(1^3F_2)$ & $2^{-+}$ & $7203$ & $T_{\psi\psi}(7300)_{\rm A'}$ & $7121$ &  & $7040$ &  & $7051$ &  \\
$T_{\psi\psi2}^\omega(1^5F_2)$ & $2^{--}$ & $7202$ &  & $7120$ &  & $7039$ &  & $7050$ &  \\
$T_{\psi\psi3}^\eta(1^3F_3)$ & $3^{-+}$ & $7210$ & $T_{\psi\psi}(7300)_{\rm A'}$ & $7128$ &  & $7047$ &  & $7057$ &  \\
$T_{\psi\psi3}^\omega(1^1F_3)$ & $3^{--}$ & $7209$ &  & $7127$ &  & $7046$ &  & $7056$ &  \\
$T_{\psi\psi3}^\omega(1^5F_3)$ & $3^{--}$ & $7208$ &  & $7126$ &  & $7045$ &  & $7055$ &  \\
$T_{\psi\psi4}^\eta(1^3F_4)$ & $4^{-+}$ & $7212$ & $T_{\psi\psi}(7300)_{\rm A'}$ & $7130$ &  & $7049$ &  & $7059$ &  \\
$T_{\psi\psi4}^\omega(1^5F_4)$ & $4^{--}$ & $7213$ &  & $7131$ &  & $7050$ &  & $7060$ &  \\
$T_{\psi\psi5}^\omega(1^5F_5)$ & $5^{--}$ & $7216$ &  & $7133$ &  & $7052$ &  & $7061$ &  \\
$T_{\psi\psi0}^f(2^5D_0)$ & $0^{++}$ & $7285$ & $T_{\psi\psi}(7300)_{\rm C}$ & $7187$ &  & $7092$ &  & $7124$ &  \\
$T_{\psi\psi1}^f(2^5D_1)$ & $1^{++}$ & $7293$ & $T_{\psi\psi}(7300)_{\rm C}$ & $7194$ & $T_{\psi\psi}(7300)_{\rm A'}$ & $7098$ &  & $7133$ &  \\
$T_{\psi\psi1}^h(2^3D_1)$ & $1^{+-}$ & $7298$ &  & $7199$ &  & $7103$ &  & $7137$ &  \\
$T_{\psi\psi2}^f(2^1D_2)$ & $2^{++}$ & $7311$ &  & $7211$ & $T_{\psi\psi}(7300)_{\rm A'}$ & $7113$ &  & $7154$ &  \\
$T_{\psi\psi2}^f(2^5D_2)$ & $2^{++}$ & $7305$ & $T_{\psi\psi}(7300)_{\rm C}$ & $7205$ & $T_{\psi\psi}(7300)_{\rm A'}$ & $7108$ &  & $7147$ &  \\
$T_{\psi\psi2}^h(2^3D_2)$ & $2^{+-}$ & $7312$ &  & $7211$ &  & $7113$ &  & $7155$ &  \\
$T_{\psi\psi3}^f(2^5D_3)$ & $3^{++}$ & $7318$ &  & $7217$ & $T_{\psi\psi}(7300)_{\rm A'}$ & $7118$ &  & $7163$ &  \\
$T_{\psi\psi3}^h(2^3D_3)$ & $3^{+-}$ & $7318$ &  & $7217$ &  & $7118$ &  & $7163$ &  \\
$T_{\psi\psi4}^f(2^5D_4)$ & $4^{++}$ & $7325$ &  & $7223$ & $T_{\psi\psi}(7300)_{\rm A'}$ & $7124$ &  & $7172$ &  \\
$T_{\psi\psi0}^\eta(3^3P_0)$ & $0^{-+}$ & $7374$ &  & $7261$ &  & $7151$ &  & $7166$ &  \\
$T_{\psi\psi1}^\eta(3^3P_1)$ & $1^{-+}$ & $7406$ &  & $7289$ & $T_{\psi\psi}(7300)_{\rm C}$ & $7175$ &  & $7239$ & $T_{\psi\psi}(7300)_{\rm A'}$ \\
$T_{\psi\psi1}^\omega(3^1P_1)$ & $1^{--}$ & $7406$ &  & $7289$ &  & $7175$ &  & $7240$ &  \\
$T_{\psi\psi1}^\omega(3^5P_1)$ & $1^{--}$ & $7376$ &  & $7262$ &  & $7151$ &  & $7173$ &  \\
$T_{\psi\psi2}^\eta(3^3P_2)$ & $2^{-+}$ & $7418$ &  & $7300$ & $T_{\psi\psi}(7300)_{\rm C}$ & $7184$ &  & $7263$ &  \\
$T_{\psi\psi2}^\omega(3^5P_2)$ & $2^{--}$ & $7415$ &  & $7297$ &  & $7181$ &  & $7256$ &  \\
$T_{\psi\psi3}^\omega(3^5P_3)$ & $3^{--}$ & $7432$ &  & $7312$ &  & $7194$ &  & $7283$ &  \\
$T_{\psi\psi0}^f(4^1S_0)$ & $0^{++}$ & $7467$ &  & $7338$ &  & $7213$ & $T_{\psi\psi}(7300)_{\rm A'}$ & $7316$ &  \\
$T_{\psi\psi1}^h(4^3S_1)$ & $1^{+-}$ & $7489$ &  & $7357$ &  & $7228$ &  & $7321$ &  \\
$T_{\psi\psi2}^f(4^5S_2)$ & $2^{++}$ & $7530$ &  & $7391$ &  & $7257$ &  & $7333$ &  \\
\bottomrule[1.0pt]\bottomrule[1.0pt]
\end{tabular}
\end{table*}

In terms of the diquark-antidiquark configuration, the total angular momentum $\bm J$ of the fully charmed tetraquark $cc\bar c\bar c$ is obtained by coupling the total angular momentum $\bm J_{cc}$ of the doubly charmed diquark $cc$, the total angular momentum $\bm J_{\bar c\bar c}$ of the doubly charmed antidiquark $\bar c\bar c$, and the relative orbital angular momentum $\bm L_\lambda$ between the diquark $cc$ and antidiquark $\bar c\bar c$, i.e.,
\begin{eqnarray}
\bm J&=&\bm J_{cc}\otimes\bm J_{\bar c\bar c}\otimes\bm L_\lambda,\label{eq31}
\end{eqnarray}
with
\begin{eqnarray}
\bm J_{cc}&=&\bm L_{cc}\otimes\bm S_{cc},\label{eq32}\\
\bm J_{\bar c\bar c}&=&\bm L_{\bar c\bar c}\otimes\bm S_{\bar c\bar c},\label{eq33}\\
\bm S_{cc}&=&\bm S_{c}\otimes\bm S_{c},\label{eq34}\\
\bm S_{\bar c\bar c}&=&\bm S_{\bar c}\otimes\bm S_{\bar c}.\label{eq35}
\end{eqnarray}
Here, $\bm S_{c}$, $\bm S_{cc}$, and $\bm L_{cc}$ ($\bm S_{\bar c}$, $\bm S_{\bar c\bar c}$, and $\bm L_{\bar c\bar c}$) denote the spin quantum number of the charm (anti)quark, the spin quantum number of the doubly charmed (anti)diquark, and the relative orbital angular momentum between the two charm (anti)quarks in the doubly charmed (anti)diquark, respectively. As mentioned previously, the spin quantum number of the color antitriplet (triplet) orbitally ground state doubly charmed diquark (antidiquark) must be $1$ due to the Pauli exclusion principle. In this work, the orbital excitations between the two (anti)quarks inside (anti)diquarks are omitted. Thereupon, the total angular momentum of the doubly charmed (anti)diquark is determined as $1$, i.e.,
\begin{eqnarray}
L_{cc}=L_{\bar c\bar c}&=&0,\label{eq36}\\
J_{cc}=S_{cc}&=&1,\label{eq37}\\
J_{\bar c\bar c}=S_{\bar c\bar c}&=&1.\label{eq38}
\end{eqnarray}

The conventional mesonic notation $n^{2S+1}L_J$ is employed to denote the fully charmed tetraquark states predicted by theories in Tables \ref{cccc1}-\ref{cccc2}. Thereinto, the principal quantum number $n$, the orbital angular momentum $L$, and the spin quantum number $\bm S$ of the fully charmed tetraquark $cc\bar c\bar c$ are
\begin{eqnarray}
n&=&n_{cc}+n_{\bar c\bar c}+n_\lambda+1,\label{eq39}\\
L&=&L_{cc}+L_{\bar c\bar c}+L_\lambda,\label{eq310}\\
\bm S&=&\bm S_{cc}\otimes\bm S_{\bar c\bar c}.\label{eq311}
\end{eqnarray}
Here, $n_{cc}$, $n_{\bar c\bar c}$, and $n_\lambda$ denote the radial quantum numbers between the two charm quarks in the diquark $cc$, between the two charm antiquarks in the antidiquark $\bar c\bar c$, and between the diquark $cc$ and antidiquark $\bar c\bar c$ in the fully charmed tetraquark $cc\bar c\bar c$, respectively. In this work, the radial excitations between the two (anti)quarks inside (anti)diquarks are omitted, thereby
\begin{eqnarray}
n&=&n_\lambda+1.\label{eq312}
\end{eqnarray}

Moreover, it is facile to espy that the total angular momentum $\bm J$ of the fully charmed tetraquark $cc\bar c\bar c$ is capable of being rewritten into the form of spin-orbit coupling \cite{Varshalovich:1988krb}, i.e.,
\begin{eqnarray}
&&\left|\left[\left(\bm L_{cc}\otimes\bm L_{\bar c\bar c}\right)_{\bm L_\rho}\otimes\bm L_\lambda\right]_{\bm L_t}\otimes\left(\bm S_{cc}\otimes\bm S_{\bar c\bar c}\right)_{\bm S}\right\rangle_{\bm J}\nonumber\\
&=&\sum_{J_\rho}\sum_{J_{cc}}\sum_{J_{\bar c\bar c}}(-1)^{L_\lambda+S+L_t+J_\rho}\sqrt{(2L_t+1)(2J_\rho+1)}\nonumber\\
&&\times\sqrt{(2L_\rho+1)(2S+1)(2J_{cc}+1)(2J_{\bar c\bar c}+1)}\nonumber\\
&&\times\left\{
\begin{array}{ccc}
L_\lambda&L_\rho&L_t\\
S&J&J_\rho
\end{array}
\right\}\left\{
\begin{array}{ccc}
L_{cc}&L_{\bar c\bar c}&L_\rho\\
S_{cc}&S_{\bar c\bar c}&S\\
J_{cc}&J_{\bar c\bar c}&J_\rho
\end{array}
\right\}\nonumber\\
&&\times\left|\left[\left(\bm L_{cc}\otimes\bm S_{cc}\right)_{\bm J_{cc}}\otimes\left(\bm L_{\bar c\bar c}\otimes\bm S_{\bar c\bar c}\right)_{\bm J_{\bar c\bar c}}\right]_{\bm J_\rho}\otimes\bm L_\lambda\right\rangle_{\bm J},\qquad\qquad\label{eq313}
\end{eqnarray}
with
\begin{eqnarray}
\bm L_t&=&\bm L_\rho\otimes\bm L_\lambda,\label{eq314}\\
\bm L_\rho&=&\bm L_{cc}\otimes\bm L_{\bar c\bar c},\label{eq315}\\
\bm J_\rho&=&\bm J_{cc}\otimes\bm J_{\bar c\bar c}.\label{eq316}
\end{eqnarray}
Apparently, the two sorts of coupling forms of the total angular momentum $\bm J$ of the tetraquark are equivalent after omitting the orbital excitations in (anti)diquarks, i.e.,
\begin{eqnarray}
\bm J&=&\bm J_\rho\otimes\bm L_\lambda=\bm S\otimes\bm L_\lambda=\bm S\otimes\bm L_t.\label{eq317}
\end{eqnarray}

As far as the tetraquark system is concerned, the internal parity $P$ is determined by the relative orbital angular momentum $L_\lambda$ between the diquark and antidiquark, the internal parity $P_{cc}$ of the diquark, and the internal parity $P_{\bar c\bar c}$ of the antidiquark \cite{Amsler:2018zkm}, i.e.,
\begin{eqnarray}
P&=&(-1)^{L_\lambda}P_{cc}P_{\bar c\bar c}=(-1)^L,\label{eq318}
\end{eqnarray}
with
\begin{eqnarray}
P_{cc}=(-1)^{L_{cc}},&&P_{\bar c\bar c}=(-1)^{L_{\bar c\bar c}}.\label{eq319}
\end{eqnarray}
Additionally, with regard to the electrically neutral flavorless self-conjugated tetraquark system with
\begin{eqnarray}
n_{cc}=n_{\bar c\bar c},\quad L_{cc}=L_{\bar c\bar c},\quad S_{cc}=S_{\bar c\bar c},\quad J_{cc}=J_{\bar c\bar c},\label{eq320}
\end{eqnarray}
the charge conjugation $C$ is determined by the relative orbital angular momentum $L_\lambda$ between the diquark and antidiquark and the total angular momentum $J_\rho$ of the diquark-antidiquark pair without interdiquark excitations \cite{Amsler:2018zkm}, i.e.,
\begin{eqnarray}
C&=&(-1)^{L_\lambda+J_\rho}.\label{eq321}
\end{eqnarray}
Whereafter, the charge conjugation $C$ of the self-conjugated tetraquark conforming to Eq. (\ref{eq320}) is capable of being expressed as
\begin{eqnarray}
C&=&(-1)^{L_\lambda+S}=(-1)^{L+S},\label{eq322}
\end{eqnarray}
after omitting the orbital excitations in (anti)diquarks. Subsequently, both the masses of the $cc\bar c\bar c$ states procured by the GI, MGI ($\mu=50$), MGI ($\mu=100$), and NR models, and the potential candidates of the newly observed $T_{\psi\psi}$ states which abide by the $J^{PC}$ constraints are enumerated in Tables \ref{cccc1}-\ref{cccc2}.

On the basis of Bose-Einstein symmetry, a pair of electrically neutral flavorless self-conjugated identical bosons, e.g., the di-photon system, are definitely the eigenstate of the charge conjugation $C$, possessing a positive charge conjugation $C$-parity \cite{Amsler:2018zkm}. Accordingly, in terms of the final state di-$J/\psi$ system, the sum of the relative orbital angular momentum $L_{\psi\psi}$ between two $J/\psi$ mesons and the spin quantum number $S_{\psi\psi}$ of two $J/\psi$ mesons must be even \cite{Zhou:2022xpd,Kuang:2023vac}, i.e.,
\begin{eqnarray}
L_{\psi\psi}+S_{\psi\psi}&=&\rm even,\label{eq323}
\end{eqnarray}
with
\begin{eqnarray}
\bm S_{\psi\psi}&=&\bm J_\psi\otimes\bm J_\psi.\label{eq324}
\end{eqnarray}
Here, $\bm J_\psi$ is the total angular momentum of the $J/\psi$ meson. Albeit the $J/\psi$+$\psi$(2S) system is not a pair of identical bosons, the charge conjugation $C$-parity is also positive for the reason that both two mesons possess the negative $C$-parities. Then, the $J^{PC}$ constraints of the fully charmed tetraquark states which are capable of decaying into the di-$J/\psi$ and $J/\psi$+$\psi$(2S) channels are illustrated in Table \ref{jpc}.

\renewcommand\tabcolsep{0.42cm}
\renewcommand{\arraystretch}{1.5}
\begin{table}[!htbp]
\caption{The $J^{PC}$ options of the di-$J/\psi$ and $J/\psi$+$\psi$(2S) channels.}\label{jpc}
\begin{tabular}{ccc}
\toprule[1.0pt]\toprule[1.0pt]
$L_{\psi\psi}$ & $J/\psi$+$J/\psi$ & $J/\psi$+$\psi$(2S) \\
\midrule[1.0pt]
$0$ & $0^{++}$, $2^{++}$ & $0^{++}$, $1^{++}$, $2^{++}$ \\
$1$ & $0^{-+}$, $1^{-+}$, $2^{-+}$ & $0^{-+}$, $1^{-+}$, $2^{-+}$, $3^{-+}$ \\
$2$ & $0^{++}$, $1^{++}$, $2^{++}$, $3^{++}$, $4^{++}$ & $0^{++}$, $1^{++}$, $2^{++}$, $3^{++}$, $4^{++}$ \\
$3$ & $2^{-+}$, $3^{-+}$, $4^{-+}$ & $1^{-+}$, $2^{-+}$, $3^{-+}$, $4^{-+}$, $5^{-+}$ \\
$\vdots$ & $\vdots$ & $\vdots$ \\
\bottomrule[1.0pt]\bottomrule[1.0pt]
\end{tabular}
\end{table}

\section{Discussion}\label{sec4}

In this section, with respect to the axial-vector doubly charmed diquark and fully charmed tetraquark states, the exhaustive comparisons on the difference of the mass spectra procured by this work (GI relativized diquark model, MGI relativized diquark model, and NR diquark model) and other phenomenological approaches are revealed. What's more, the potential interpretations for the newly observed $T_{\psi\psi}$ states are performed as well.

\renewcommand\tabcolsep{0.29cm}
\renewcommand{\arraystretch}{1.5}
\begin{table}[!htbp]
\caption{A comparison of the ground state axial-vector doubly charmed diquark masses from this work (GI, MGI, and NR models) and other phenomenological approaches (in unit of MeV).}\label{cc}
\begin{tabular}{cccc}
\toprule[1.0pt]\toprule[1.0pt]
Reference & Mass & Reference & Mass \\
\midrule[1.0pt]
\cite{Tang:2011fv} & $2770$ & \cite{Martynenko:2013eoa} I & $3233$ \\
\cite{Song:2022csw} & $2865.5$ & \cite{Giannuzzi:2009gh} & $3238$ \\
\cite{Yu:2006ty} III & $2876$ & \cite{Yu:2006ty} IX & $3247$ \\
\cite{Sonnenschein:2020nwn} & $2980$ & \cite{Pal:2021gkr} & $3270.5$ \\
\cite{Tang:2012np} & $2990\pm100$ & \cite{Lu:2017meb} & $3294$ \\
\cite{Zhu:2020xni} & $3000$ & \cite{Yin:2019bxe} & $3300$ \\
\cite{Yu:2006ty} VI & $3090$ & MGI ($\mu=100$) & $3300$ \\
\cite{Mutuk:2021hmi} & $3114$ & \cite{Li:2021ygk} I & $3303$ \\
\cite{Gutierrez-Guerrero:2019uwa} I & $3120$ & \cite{Nefediev:2021pww} & $3310$ \\
\cite{Tiwari:2021tmz} & $3124$ & MGI ($\mu=50$) & $3314$ \\
\cite{Lundhammar:2020xvw} & $3128$ & \cite{Martynenko:2013eoa} II & $3328$ \\
\cite{Kiselev:2002iy} & $3130$ & GI \cite{Ferretti:2019zyh} & $3329$ \\
\cite{Mutuk:2021epz} & $3133$ & \cite{Yu:2018com} II & $3370$ \\
\cite{Debastiani:2017msn} & $3133.4$ & \cite{Yu:2018com} IV & $3370$ \\
\cite{Li:2021ygk} II & $3135$ & \cite{Yu:2006ty} V & $3371$ \\
\cite{Chao:1980dv} & $3136.4$ & \cite{Yu:2006ty} XII & $3381$ \\
\cite{Giron:2020wpx} & $3136.4\pm10$ & \cite{Yu:2018com} I & $3400$ \\
\cite{Bedolla:2019noq} & $3144$ & \cite{Yu:2018com} III & $3420$ \\
NR & $3152$ & \cite{Eichmann:2020oqt} & $3423\pm8$ \\
\cite{Pal:2023yid} & $3153.1$ & \cite{Yu:2006ty} I & $3460$ \\
\cite{Gershtein:2000nx} & $3160$ & \cite{Yu:2006ty} XV & $3500$ \\
\cite{Yu:2006ty} II & $3169$ & \cite{Esau:2019hqw} & $3510\pm350$ \\
\cite{Weng:2020jao} & $3171.51$ & \cite{Yu:2006ty} VIII & $3520$ \\
\cite{Guo:2021yws} & $3182.67\pm30$ & \cite{Yu:2006ty} IV & $3648$ \\
\cite{Karliner:2016zzc} & $3204.1$ & \cite{Yu:2006ty} XI & $3648$ \\
\cite{Gutierrez-Guerrero:2019uwa} II & $3210$ & \cite{Yu:2006ty} XIV & $3760$ \\
\cite{Gutierrez-Guerrero:2021rsx} & $3220$ & \cite{Yu:2006ty} VII & $3789$ \\
\cite{Martynenko:2014ola} & $3224$ & \cite{Yu:2006ty} X & $3909$ \\
\cite{Faustov:2020qfm} & $3226$ & \cite{Yu:2006ty} XIII & $4015$ \\
\bottomrule[1.0pt]\bottomrule[1.0pt]
\end{tabular}
\end{table}

\subsection{Comparison on the $cc$ diquark}\label{subsec41}

As far as the color antitriplet ground state axial-vector doubly charmed diquark mass is concerned, there are conspicuous discrepancies among the multifarious theoretical approaches \cite{Chao:1980dv,Lu:2017meb,Karliner:2016zzc,Debastiani:2017msn,Esau:2019hqw,Faustov:2020qfm,Lundhammar:2020xvw,Giron:2020wpx,Sonnenschein:2020nwn,Zhu:2020xni,
Pal:2021gkr,Li:2021ygk,Mutuk:2021hmi,Weng:2020jao,Nefediev:2021pww,Tiwari:2021tmz,Pal:2023yid,Ferretti:2019zyh,Gershtein:2000nx,Kiselev:2002iy,Yu:2006ty,
Giannuzzi:2009gh,Tang:2012np,Tang:2011fv,Martynenko:2013eoa,Martynenko:2014ola,Yu:2018com,Yin:2019bxe,Gutierrez-Guerrero:2019uwa,Bedolla:2019noq,Eichmann:2020oqt,
Gutierrez-Guerrero:2021rsx,Guo:2021yws,Mutuk:2021epz,Song:2022csw}. As Table \ref{cc} lays out, the multitudinous theoretical values of the doubly charmed diquark mass staggeringly lie on the extensive domain between 2770 \cite{Tang:2011fv} and 4015 MeV \cite{Yu:2006ty}. Evidently, the doubly charmed diquark mass 3329 MeV from the GI model in this work is completely coherent with the result in Ref. \cite{Ferretti:2019zyh}. Additionally, the theoretical value 3329 MeV of the GI model is very close to the doubly charmed diquark mass 3328 MeV acquired by Ref. \cite{Martynenko:2013eoa}. The theoretical values 3314 and 3300 MeV of the MGI model are very close to the doubly charmed diquark masses 3310, 3303, 3300, and 3294 MeV adopted by Refs. \cite{Nefediev:2021pww}, \cite{Li:2021ygk}, \cite{Yin:2019bxe}, and \cite{Lu:2017meb}, respectively. The theoretical value 3152 MeV of the NR model is very close to the doubly charmed diquark masses 3144, 3153, and 3160 MeV procured by Refs. \cite{Bedolla:2019noq}, \cite{Pal:2023yid}, and \cite{Gershtein:2000nx}, respectively.

\renewcommand\tabcolsep{0.055cm}
\renewcommand{\arraystretch}{1.5}
\begin{table*}[!htbp]
\caption{A comparison of the $1S$-wave fully charmed tetraquark masses from this work (GI, MGI, and NR models) and other phenomenological approaches (in unit of MeV).}\label{T4c1}
\begin{tabular}{cccccccccccc}
\toprule[1.0pt]\toprule[1.0pt]
Reference & $1^1S_0$ & $1^3S_1$ & $1^5S_2$ & Reference & $1^1S_0$ & $1^3S_1$ & $1^5S_2$ & Reference & $1^1S_0$ & $1^3S_1$ & $1^5S_2$ \\
\midrule[1.0pt]
\cite{Heupel:2012ua} & $5300\pm500$ & $\cdots$ & $\cdots$ & \cite{Nefediev:2021pww} & $6196$ & $\cdots$ & $6560$ & \cite{Yang:2020atz} & $6469$ & $6674$ & $7026$ \\
\cite{Bedolla:2019zwg} & $5883$ & $6120$ & $6246$ & \cite{Lundhammar:2020xvw} II & $6198$ & $6246$ & $6323$ & \cite{Albuquerque:2020hio} VI & $6470\pm70$ & $\cdots$ & $\cdots$ \\
\cite{Tiwari:2021tmz} & $5942$ & $5989$ & $6082$ & \cite{Iwasaki:1975pv} & $6200$ & $\cdots$ & $\cdots$ & \cite{Zhao:2020nwy} & $6476$ & $6441$ & $6475$ \\
NR & $5944$ & $6001$ & $6105$ & \cite{Wang:2022xja} & $6200\pm100$ & $6240\pm100$ & $6270\pm90$ & \cite{Lloyd:2003yc} II & $6477$ & $6528$ & $6573$ \\
\cite{Lundhammar:2020xvw} I & $5960$ & $6009$ & $6100$ & \cite{Ke:2021iyh} & $6201$ & $6396$ & $6391$ & \cite{Jin:2020jfc} II & $6482$ & $6488$ & $6499$ \\
\cite{Berezhnoy:2011xn} & $5966$ & $6051$ & $6223$ & \cite{Giron:2020wpx} & $6265.05\pm1.05$ & $\cdots$ & $\cdots$ & \cite{Wang:2019rdo} II & $6483$ & $6450$ & $6479$ \\
\cite{Debastiani:2017msn} & $5969.4$ & $6020.9$ & $6115.4$ & \cite{Jin:2020jfc} V & $6270$ & $6285$ & $6314$ & \cite{Liu:2019zuc} & $6487$ & $6500$ & $6524$ \\
\cite{Majarshin:2021hex} & $5978$ & $6155$ & $6263$ & \cite{Weng:2020jao} & $6271.3$ & $6230.6$ & $6287.3$ & \cite{Badalian:1985es} II & $6490$ & $\cdots$ & $\cdots$ \\
MGI ($\mu=100$) & $5989$ & $6115$ & $6260$ & \cite{Ader:1981db} III & $6276$ & $\cdots$ & $\cdots$ & \cite{Albuquerque:2020hio} II & $6490\pm70$ & $\cdots$ & $\cdots$ \\
\cite{Wang:2017jtz} & $5990\pm80$ & $\cdots$ & $6090\pm80$ & \cite{Mutuk:2021hmi} & $6322$ & $6354$ & $6385$ & \cite{Jin:2020jfc} I & $6493$ & $6495$ & $6498$ \\
\cite{Wang:2018poa} & $\cdots$ & $6050\pm80$ & $\cdots$ & \cite{Gordillo:2020sgc} & $6351$ & $6441$ & $6471$ & \cite{Wang:2021kfv} & $6498$ & $6481$ & $6502$ \\
\cite{Wu:2016vtq} IV & $6001$ & $6109$ & $6166$ & \cite{Jin:2020jfc} VI & $6358$ & $6375$ & $6407$ & \cite{Albuquerque:2020hio} IV & $6500$ & $\cdots$ & $\cdots$ \\
\cite{Heller:1985cb} & $6011$ & $\cdots$ & $\cdots$ & \cite{Chen:2021crg} & $6360.2$ & $6397.6$ & $6410.4$ & \cite{Lu:2020cns} & $6501$ & $6515$ & $6543$ \\
MGI ($\mu=50$) & $6020$ & $6148$ & $6295$ & \cite{Lloyd:2003yc} I & $6367$ & $6411$ & $6459$ & \cite{Yu:2022lak} & $6503$ & $6517$ & $6544$ \\
\cite{Wu:2016vtq} III & $6022$ & $6126$ & $6183$ & \cite{An:2022qpt} & $6384.4$ & $6451.5$ & $6482.7$ & \cite{Badalian:1985es} I & $6510$ & $\cdots$ & $\cdots$ \\
\cite{Wu:2016vtq} II & $6035$ & $6137$ & $6194$ & \cite{Zhang:2022qtp} II & $6411$ & $6453$ & $6475$ & \cite{Albuquerque:2020hio} V & $6510\pm60$ & $\cdots$ & $\cdots$ \\
\cite{Barnea:2006sd} II & $6038$ & $6101$ & $6172$ & \cite{Zhang:2022qtp} I & $6414$ & $6414$ & $6414$ & \cite{Albuquerque:2020hio} I & $6520$ & $\cdots$ & $\cdots$ \\
GI & $6053$ & $6181$ & $6331$ & \cite{Li:2021ygk} & $6419$ & $6456$ & $6516$ & \cite{Deng:2020iqw} II & $6573$ & $6580$ & $6607$ \\
\cite{Zhu:2020xni} & $6055^{+69}_{-74}$ & $\cdots$ & $6090^{+62}_{-66}$ & \cite{Yang:2021hrb} & $6421$ & $6439$ & $6472$ & \cite{Wu:2022qwd} II & $6600^{+90}_{-100}$ & $\cdots$ & $6980^{+90}_{-110}$ \\
\cite{Wu:2022qwd} I & $6070^{+50}_{-70}$ & $\cdots$ & $6070^{+80}_{-100}$ & \cite{Wang:2019rdo} I & $6425$ & $6425$ & $6432$ & \cite{Wu:2016vtq} I & $6797$ & $6899$ & $6956$ \\
\cite{Deng:2020iqw} I & $6109$ & $6139$ & $6194$ & \cite{Ader:1981db} I & $6437$ & $\cdots$ & $\cdots$ & \cite{Asadi:2021ids} III & $6821.7$ & $6821.8$ & $6821.8$ \\
\cite{Zhuang:2021pci} & $6109.05$ & $6137.30$ & $6193.80$ & \cite{Ader:1981db} II & $6450$ & $\cdots$ & $\cdots$ & \cite{Asadi:2021ids} I & $6850.0$ & $6870.4$ & $6912.7$ \\
\cite{Barnea:2006sd} I & $6115$ & $6176$ & $6216$ & \cite{Albuquerque:2020hio} III & $6450\pm80$ & $\cdots$ & $\cdots$ & \cite{Asadi:2021ids} II & $6874.2$ & $6913.3$ & $6989.9$ \\
\cite{Jin:2020jfc} IV & $6128$ & $6149$ & $6197$ & \cite{Deng:2020iqw} III & $6454$ & $6463$ & $6486$ & \cite{Sonnenschein:2020nwn} & $6895$ & $\cdots$ & $\cdots$ \\
\cite{Anwar:2017toa} & $6140$ & $\cdots$ & $\cdots$ & \cite{Chen:2016jxd} & $6460\pm160$ & $\cdots$ & $6510\pm150$ & \cite{Esau:2019hqw} & $7000$ & $7000$ & $7000$ \\
\cite{Faustov:2020qfm} & $6190$ & $6271$ & $6367$ & \cite{Zhang:2020xtb} & $6460^{+130}_{-170}$ & $\cdots$ & $\cdots$ & \cite{Wang:2022yes} & $7035.1$ & $7049.6$ & $7068.5$ \\
\cite{Karliner:2016zzc} & $6191.5\pm25$ & $\cdots$ & $\cdots$ & \cite{Jin:2020jfc} III & $6466$ & $6479$ & $6505$ & \cite{Pal:2021gkr} II & $\cdots$ & $\cdots$ & $7295.0$ \\
\cite{Karliner:2020dta} & $\cdots$ & $\cdots$ & $6429\pm25$ & \cite{Zhao:2020zjh} & $6466$ & $6494$ & $6551$ & \cite{Kuang:2022vdy} & $7438\pm2$ & $\cdots$ & $\cdots$ \\
\cite{Pal:2023yid} & $\cdots$ & $\cdots$ & $6520.4$ & \cite{Pal:2021gkr} I & $\cdots$ & $\cdots$ & $6888.4$ & $\cdots$ & $\cdots$ & $\cdots$ & $\cdots$ \\
\bottomrule[1.0pt]\bottomrule[1.0pt]
\end{tabular}
\end{table*}

\renewcommand\tabcolsep{0.045cm}
\renewcommand{\arraystretch}{1.5}
\begin{table*}[!htbp]
\caption{A comparison of the $2S$-, $3S$-, and $4S$-wave fully charmed tetraquark masses from this work (GI, MGI, and NR models) and other phenomenological approaches (in unit of MeV).}\label{T4c2}
\begin{tabular}{cccccccccccc}
\toprule[1.0pt]\toprule[1.0pt]
Reference & $2^1S_0$ & $2^3S_1$ & $2^5S_2$ & Reference & $2^1S_0$ & $2^3S_1$ & $2^5S_2$ & Reference & $2^1S_0$ & $2^3S_1$ & $2^5S_2$ \\
\midrule[1.0pt]
\cite{Wang:2020ols} & $6480\pm80$ & $6520\pm80$ & $6560\pm80$ & MGI ($\mu=50$) & $6697$ & $6738$ & $6803$ & \cite{Zhao:2020nwy} & $6908$ & $6896$ & $6921$ \\
\cite{Zhu:2020xni} & $6555^{+36}_{-37}$ & $\cdots$ & $6566^{+34}_{-35}$ & \cite{Lloyd:2003yc} I & $6719$ & $\cdots$ & $\cdots$ & \cite{Jin:2020jfc} I & $6910$ & $6740$ & $6725$ \\
\cite{Wang:2022xja} & $6570\pm90$ & $6640\pm90$ & $6690\pm90$ & GI & $6751$ & $6794$ & $6864$ & \cite{Lu:2020cns} & $6917$ & $6928$ & $6948$ \\
\cite{Bedolla:2019zwg} & $6573$ & $6669$ & $6739$ & \cite{Giron:2020wpx} & $6771.8$ & $\cdots$ & $\cdots$ & \cite{Liu:2021rtn} & $6930$ & $6934$ & $6942$ \\
\cite{Mutuk:2021hmi} & $6575$ & $6609$ & $6639$ & \cite{Faustov:2021hjs} & $6782$ & $6816$ & $6868$ & \cite{Yu:2022lak} & $6937$ & $6932$ & $6952$ \\
\cite{Ke:2021iyh} & $6575$ & $6799$ & $6794$ & \cite{Jin:2020jfc} II & $6825$ & $6830$ & $6825$ & \cite{Jin:2020jfc} V & $6950$ & $6925$ & $6900$ \\
\cite{Tiwari:2021tmz} & $6644$ & $6656$ & $6678$ & \cite{Wang:2019rdo} I & $\cdots$ & $6856$ & $6864$ & \cite{Jin:2020jfc} IV & $6950$ & $7040$ & $7010$ \\
MGI ($\mu=100$) & $6644$ & $6683$ & $6744$ & \cite{Wang:2019rdo} II & $\cdots$ & $6894$ & $6919$ & \cite{Jin:2020jfc} VI & $6975$ & $7150$ & $7050$ \\
\cite{Debastiani:2017msn} & $6663.3$ & $6674.5$ & $6698.1$ & \cite{Karliner:2020dta} & $6871\pm25$ & $\cdots$ & $6967\pm25$ & \cite{Wang:2021kfv} & $7007$ & $6954$ & $6917$ \\
NR & $6667$ & $6679$ & $6703$ & \cite{Zhao:2020zjh} & $6883$ & $6911$ & $6968$ & \cite{Sonnenschein:2020nwn} & $7185\pm45$ & $\cdots$ & $\cdots$ \\
\cite{Lloyd:2003yc} II & $6695$ & $\cdots$ & $\cdots$ & \cite{Jin:2020jfc} III & $6900$ & $6960$ & $6970$ & \cite{Wang:2022yes} & $7202.2$ & $7273.5$ & $7281.3$ \\
\midrule[1.0pt]
Reference & $3^1S_0$ & $3^3S_1$ & $3^5S_2$ & Reference & $3^1S_0$ & $3^3S_1$ & $3^5S_2$ & Reference & $3^1S_0$ & $3^3S_1$ & $3^5S_2$ \\
\midrule[1.0pt]
\cite{Mutuk:2021hmi} & $6782$ & $6814$ & $6842$ & MGI ($\mu=100$) & $6979$ & $7001$ & $7040$ & \cite{Liu:2021rtn} & $7241$ & $7243$ & $7248$ \\
\cite{Zhu:2020xni} & $6883\pm27$ & $\cdots$ & $6890^{+27}_{-26}$ & \cite{Tiwari:2021tmz} & $7011$ & $7018$ & $7033$ & \cite{Jin:2020jfc} VI & $7250$ & $7250$ & $\cdots$ \\
\cite{Wang:2019rdo} I & $\cdots$ & $6915$ & $6919$ & NR & $7031$ & $7038$ & $7054$ & \cite{Jin:2020jfc} I & $7250$ & $7275$ & $\cdots$ \\
\cite{Wang:2021kfv} & $\cdots$ & $7024$ & $7030$ & \cite{Lu:2020cns} & $7046$ & $7052$ & $7064$ & \cite{Jin:2020jfc} IV & $7250$ & $7280$ & $\cdots$ \\
\cite{Wang:2019rdo} II & $\cdots$ & $7036$ & $7058$ & MGI ($\mu=50$) & $7064$ & $7089$ & $7134$ & \cite{Faustov:2021hjs} & $7259$ & $7287$ & $7333$ \\
\cite{Ke:2021iyh} & $6897$ & $7148$ & $7148$ & GI & $7152$ & $7180$ & $7230$ & \cite{Jin:2020jfc} III & $7260$ & $7250$ & $\cdots$ \\
\cite{Wang:2022xja} & $6920\pm90$ & $7030\pm90$ & $7090\pm90$ & \cite{Jin:2020jfc} II & $7210$ & $7280$ & $\cdots$ & \cite{Zhao:2020nwy} & $7296$ & $7300$ & $7320$ \\
\cite{Wang:2020ols} & $6940\pm80$ & $6960\pm80$ & $7000\pm80$ & \cite{Jin:2020jfc} V & $7225$ & $7250$ & $\cdots$ & \cite{Sonnenschein:2020nwn} & $7440\pm90$ & $\cdots$ & $\cdots$ \\
\cite{Bedolla:2019zwg} & $6948$ & $7016$ & $7071$ & \cite{Zhao:2020zjh} & $7225$ & $7253$ & $7310$ & $\cdots$ & $\cdots$ & $\cdots$ & $\cdots$ \\
\midrule[1.0pt]
Reference & $4^1S_0$ & $4^3S_1$ & $4^5S_2$ & Reference & $4^1S_0$ & $4^3S_1$ & $4^5S_2$ & Reference & $4^1S_0$ & $4^3S_1$ & $4^5S_2$ \\
\midrule[1.0pt]
\cite{Zhu:2020xni} & $7154\pm22$ & $\cdots$ & $7160^{+21}_{-22}$ & \cite{Wang:2022xja} & $7250\pm90$ & $7400\pm90$ & $7460\pm90$ & \cite{Wang:2020ols} & $7360\pm80$ & $7370\pm80$ & $7410\pm80$ \\
MGI ($\mu=100$) & $7213$ & $7228$ & $7257$ & NR & $7316$ & $7321$ & $7333$ & GI & $7467$ & $7489$ & $7530$ \\
\cite{Bedolla:2019zwg} & $7237$ & $7293$ & $\cdots$ & MGI ($\mu=50$) & $7338$ & $7357$ & $7391$ & \cite{Sonnenschein:2020nwn} & $7680\pm120$ & $\cdots$ & $\cdots$ \\
\bottomrule[1.0pt]\bottomrule[1.0pt]
\end{tabular}
\end{table*}

\renewcommand\tabcolsep{0.49cm}
\renewcommand{\arraystretch}{1.5}
\begin{table*}[!htbp]
\caption{A comparison of the $1P$- and $2P$-wave fully charmed tetraquark masses from this work (GI, MGI, and NR models) and other phenomenological approaches (in unit of MeV).}\label{T4c3}
\begin{tabular}{cccccccc}
\toprule[1.0pt]\toprule[1.0pt]
Reference & $1^3P_0$ & $1^3P_1$ & $1^1P_1$ & $1^5P_1$ & $1^3P_2$ & $1^5P_2$ & $1^5P_3$ \\
\midrule[1.0pt]
\cite{Chao:1980dv} & $\cdots$ & $\cdots$ & $6550$ & $6390$ & $\cdots$ & $\cdots$ & $\cdots$ \\
\cite{Tiwari:2021tmz} & $6462$ & $6556$ & $6555$ & $6461$ & $6589$ & $6579$ & $6625$ \\
\cite{Debastiani:2017msn} & $6480.4$ & $6577.4$ & $6577.1$ & $6495.4$ & $6609.9$ & $6600.2$ & $6641.2$ \\
\cite{Bedolla:2019zwg} & $6596$ & $\cdots$ & $6580$ & $6584$ & $\cdots$ & $\cdots$ & $\cdots$ \\
NR & $6478$ & $6584$ & $6584$ & $6495$ & $6618$ & $6609$ & $6648$ \\
MGI ($\mu=100$) & $6545$ & $6604$ & $6605$ & $6544$ & $6623$ & $6618$ & $6643$ \\
\cite{Faustov:2021hjs} & $6628$ & $6634$ & $6631$ & $6635$ & $6644$ & $6648$ & $6664$ \\
MGI ($\mu=50$) & $6588$ & $6650$ & $6651$ & $6588$ & $6670$ & $6664$ & $6691$ \\
GI & $6633$ & $6697$ & $6698$ & $6633$ & $6718$ & $6712$ & $6739$ \\
\cite{Ader:1981db} II & $\cdots$ & $\cdots$ & $6714$ & $\cdots$ & $\cdots$ & $\cdots$ & $\cdots$ \\
\cite{Ader:1981db} I & $\cdots$ & $\cdots$ & $6718$ & $\cdots$ & $\cdots$ & $\cdots$ & $\cdots$ \\
\cite{Giron:2020wpx} II & $6597.19$ & $6691.79$ & $6726.68$ & $6556.22$ & $6771.55$ & $6687.87$ & $6817.51$ \\
\cite{Giron:2020wpx} I & $6595.79$ & $6704.69$ & $6727.98$ & $6563.70$ & $6764.09$ & $6713.49$ & $6802.59$ \\
\cite{Deng:2020iqw} III & $\cdots$ & $\cdots$ & $6730$ & $\cdots$ & $6735$ & $\cdots$ & $6744$ \\
\cite{Badalian:1985es} II & $\cdots$ & $\cdots$ & $6740$ & $\cdots$ & $\cdots$ & $\cdots$ & $\cdots$ \\
\cite{Wang:2021kfv} & $6726$ & $6743$ & $6740$ & $6718$ & $6752$ & $6739$ & $6752$ \\
\cite{Badalian:1985es} I & $\cdots$ & $\cdots$ & $6770$ & $6750$ & $\cdots$ & $\cdots$ & $\cdots$ \\
\cite{liu:2020eha} & $6750$ & $6769$ & $6770$ & $6754$ & $6783$ & $6781$ & $6801$ \\
\cite{Yu:2022lak} & $6796$ & $6796$ & $6791$ & $6808$ & $6797$ & $6808$ & $6809$ \\
\cite{Deng:2020iqw} II & $\cdots$ & $\cdots$ & $6901$ & $\cdots$ & $6912$ & $\cdots$ & $6924$ \\
\cite{Lloyd:2003yc} I & $6876$ & $6906$ & $6917$ & $\cdots$ & $6926$ & $\cdots$ & $\cdots$ \\
\cite{Lloyd:2003yc} II & $6969$ & $7004$ & $7013$ & $\cdots$ & $7033$ & $\cdots$ & $\cdots$ \\
\cite{Sonnenschein:2020nwn} & $\cdots$ & $\cdots$ & $7110\pm50$ & $\cdots$ & $\cdots$ & $\cdots$ & $\cdots$ \\
\midrule[1.0pt]
Reference & $2^3P_0$ & $2^3P_1$ & $2^1P_1$ & $2^5P_1$ & $2^3P_2$ & $2^5P_2$ & $2^5P_3$ \\
\midrule[1.0pt]
\cite{Tiwari:2021tmz} & $6852$ & $6927$ & $6926$ & $6850$ & $6952$ & $6945$ & $6983$ \\
MGI ($\mu=100$) & $6902$ & $6937$ & $6937$ & $6902$ & $6949$ & $6946$ & $6963$ \\
\cite{Bedolla:2019zwg} & $6953$ & $\cdots$ & $6940$ & $6943$ & $\cdots$ & $\cdots$ & $\cdots$ \\
\cite{Debastiani:2017msn} & $6866.5$ & $6943.9$ & $6944.1$ & $6875.6$ & $6970.4$ & $6962.1$ & $6996.7$ \\
NR & $6867$ & $6951$ & $6951$ & $6877$ & $6977$ & $6970$ & $7002$ \\
\cite{Giron:2020wpx} & $\cdots$ & $\cdots$ & $7011.9\pm1.1$ & $\cdots$ & $\cdots$ & $\cdots$ & $\cdots$ \\
MGI ($\mu=50$) & $6975$ & $7013$ & $7013$ & $6975$ & $7027$ & $7023$ & $7041$ \\
\cite{Faustov:2021hjs} & $7100$ & $7099$ & $7091$ & $7113$ & $7098$ & $7113$ & $7112$ \\
GI & $7050$ & $7091$ & $7091$ & $7051$ & $7106$ & $7102$ & $7122$ \\
\cite{Yu:2022lak} & $7146$ & $7146$ & $7142$ & $7155$ & $7147$ & $7155$ & $7156$ \\
\bottomrule[1.0pt]\bottomrule[1.0pt]
\end{tabular}
\end{table*}

\renewcommand\tabcolsep{0.41cm}
\renewcommand{\arraystretch}{1.5}
\begin{table*}[!htbp]
\caption{A comparison of the $3P$-, $1D$-, $2D$-, and $1F$-wave fully charmed tetraquark masses from this work (GI, MGI, and NR models) and other phenomenological approaches (in unit of MeV).}\label{T4c4}
\begin{tabular}{cccccccccc}
\toprule[1.0pt]\toprule[1.0pt]
Reference & $\cdots$ & $3^3P_0$ & $3^3P_1$ & $3^1P_1$ & $3^5P_1$ & $3^3P_2$ & $3^5P_2$ & $3^5P_3$ & $\cdots$ \\
\midrule[1.0pt]
MGI ($\mu=100$) & $\cdots$ & $7151$ & $7175$ & $7175$ & $7151$ & $7184$ & $7181$ & $7194$ & $\cdots$ \\
\cite{Tiwari:2021tmz} & $\cdots$ & $7154$ & $7221$ & $7222$ & $7151$ & $7244$ & $7237$ & $7272$ & $\cdots$ \\
\cite{Bedolla:2019zwg} & $\cdots$ & $7236$ & $\cdots$ & $7226$ & $7229$ & $\cdots$ & $\cdots$ & $\cdots$ & $\cdots$ \\
NR & $\cdots$ & $7166$ & $7239$ & $7240$ & $7173$ & $7263$ & $7256$ & $7283$ & $\cdots$ \\
MGI ($\mu=50$) & $\cdots$ & $7261$ & $7289$ & $7289$ & $7262$ & $7300$ & $7297$ & $7312$ & $\cdots$ \\
GI & $\cdots$ & $7374$ & $7406$ & $7406$ & $7376$ & $7418$ & $7415$ & $7432$ & $\cdots$ \\
\midrule[1.0pt]
Reference & $1^5D_0$ & $1^5D_1$ & $1^3D_1$ & $1^1D_2$ & $1^5D_2$ & $1^3D_2$ & $1^5D_3$ & $1^3D_3$ & $1^5D_4$ \\
\midrule[1.0pt]
\cite{Chao:1980dv} & $\cdots$ & $\cdots$ & $\cdots$ & $6780$ & $\cdots$ & $\cdots$ & $\cdots$ & $\cdots$ & $\cdots$ \\
\cite{Bedolla:2019zwg} & $6827$ & $6827$ & $6829$ & $6827$ & $6827$ & $\cdots$ & $\cdots$ & $\cdots$ & $\cdots$ \\
NR & $6826$ & $6835$ & $6841$ & $6859$ & $6850$ & $6860$ & $6867$ & $6867$ & $6876$ \\
MGI ($\mu=100$) & $6831$ & $6839$ & $6846$ & $6860$ & $6853$ & $6860$ & $6867$ & $6867$ & $6875$ \\
\cite{Giron:2020wpx} & $\cdots$ & $\cdots$ & $\cdots$ & $6861.45\pm0.95$ & $\cdots$ & $\cdots$ & $\cdots$ & $\cdots$ & $\cdots$ \\
\cite{Faustov:2021hjs} & $6899$ & $6904$ & $6909$ & $6921$ & $6915$ & $6920$ & $6929$ & $6932$ & $6945$ \\
MGI ($\mu=50$) & $6891$ & $6900$ & $6907$ & $6922$ & $6915$ & $6923$ & $6929$ & $6930$ & $6938$ \\
GI & $6953$ & $6963$ & $6970$ & $6985$ & $6978$ & $6986$ & $6994$ & $6994$ & $7002$ \\
\cite{Liu:2021rtn} & $6968$ & $6976$ & $6978$ & $6986$ & $6989$ & $6985$ & $7005$ & $7002$ & $7020$ \\
\cite{Deng:2020iqw} III & $\cdots$ & $\cdots$ & $\cdots$ & $6995$ & $\cdots$ & $\cdots$ & $\cdots$ & $6947$ & $6951$ \\
\cite{Deng:2020iqw} II & $\cdots$ & $\cdots$ & $\cdots$ & $7182$ & $\cdots$ & $\cdots$ & $\cdots$ & $7185$ & $7191$ \\
\cite{Sonnenschein:2020nwn} & $\cdots$ & $\cdots$ & $\cdots$ & $7305\pm85$ & $\cdots$ & $\cdots$ & $\cdots$ & $\cdots$ & $\cdots$ \\
\midrule[1.0pt]
Reference & $2^5D_0$ & $2^5D_1$ & $2^3D_1$ & $2^1D_2$ & $2^5D_2$ & $2^3D_2$ & $2^5D_3$ & $2^3D_3$ & $2^5D_4$ \\
\midrule[1.0pt]
MGI ($\mu=100$) & $7092$ & $7098$ & $7103$ & $7113$ & $7108$ & $7113$ & $7118$ & $7118$ & $7124$ \\
\cite{Bedolla:2019zwg} & $7125$ & $7125$ & $7128$ & $7126$ & $7125$ & $\cdots$ & $\cdots$ & $\cdots$ & $\cdots$ \\
NR & $7124$ & $7133$ & $7137$ & $7154$ & $7147$ & $7155$ & $7163$ & $7163$ & $7172$ \\
MGI ($\mu=50$) & $7187$ & $7194$ & $7199$ & $7211$ & $7205$ & $7211$ & $7217$ & $7217$ & $7223$ \\
\cite{Giron:2020wpx} & $\cdots$ & $\cdots$ & $\cdots$ & $7215.0\pm1.7$ & $\cdots$ & $\cdots$ & $\cdots$ & $\cdots$ & $\cdots$ \\
GI & $7285$ & $7293$ & $7298$ & $7311$ & $7305$ & $7312$ & $7318$ & $7318$ & $7325$ \\
\midrule[1.0pt]
Reference & $1^5F_1$ & $1^3F_2$ & $1^5F_2$ & $1^3F_3$ & $1^1F_3$ & $1^5F_3$ & $1^3F_4$ & $1^5F_4$ & $1^5F_5$ \\
\midrule[1.0pt]
\cite{Chao:1980dv} & $\cdots$ & $\cdots$ & $\cdots$ & $\cdots$ & $6980$ & $\cdots$ & $\cdots$ & $\cdots$ & $\cdots$ \\
MGI ($\mu=100$) & $7035$ & $7040$ & $7039$ & $7047$ & $7046$ & $7045$ & $7049$ & $7050$ & $7052$ \\
NR & $7046$ & $7051$ & $7050$ & $7057$ & $7056$ & $7055$ & $7059$ & $7060$ & $7061$ \\
MGI ($\mu=50$) & $7115$ & $7121$ & $7120$ & $7128$ & $7127$ & $7126$ & $7130$ & $7131$ & $7133$ \\
GI & $7197$ & $7203$ & $7202$ & $7210$ & $7209$ & $7208$ & $7212$ & $7213$ & $7216$ \\
\cite{Sonnenschein:2020nwn} & $\cdots$ & $\cdots$ & $\cdots$ & $\cdots$ & $7485\pm125$ & $\cdots$ & $\cdots$ & $\cdots$ & $\cdots$ \\
\bottomrule[1.0pt]\bottomrule[1.0pt]
\end{tabular}
\end{table*}

\subsection{Comparison on the $cc\bar c\bar c$ tetraquark}\label{subsec42}

The comparison and discussion on the difference of the fully charmed tetraquark family spectroscopy, acquired by this work (GI, MGI, and NR models) and other phenomenological approaches, are specifically exhibited as follows.

\subsubsection{$1S$-wave}\label{subsec421}

Indeed, the ground state $1S$-wave fully charmed tetraquark states have been predicted by phenomenological theories for the most times in the entire $cc\bar c\bar c$ family, as Table \ref{T4c1} demonstrates. Nevertheless, the theoretical predictions of the $1S$-wave $cc\bar c\bar c$ states are tremendously discrepant, distributing throughout the gigantic interval of practically 2 GeV between 5300 \cite{Heupel:2012ua} and 7438 MeV \cite{Kuang:2022vdy}. It embodies the existence of huge divarication in terms of the status quo of the phenomenological studies on the spectroscopic properties of the fully charmed tetraquark states. In this work, the predicted values of the $1S$-wave $cc\bar c\bar c$ states acquired by the GI, MGI, and NR models lie on the energy sector between 5944 and 6331 MeV, jibing with the corresponding results given by the earliest theoretical estimation \cite{Iwasaki:1975pv}, the quark bag model \cite{Heller:1985cb,Ader:1981db}, the quark potential model \cite{Bedolla:2019zwg,Tiwari:2021tmz,Lundhammar:2020xvw,Berezhnoy:2011xn,Debastiani:2017msn,Wu:2016vtq,Barnea:2006sd,Deng:2020iqw,Zhuang:2021pci,Jin:2020jfc,
Anwar:2017toa,Faustov:2020qfm,Karliner:2016zzc,Nefediev:2021pww,Weng:2020jao,Mutuk:2021hmi}, the Bethe-Salpeter (BS) equation \cite{Zhu:2020xni,Ke:2021iyh}, the QCD sum rules \cite{Wang:2017jtz,Wang:2018poa,Wu:2022qwd,Wang:2022xja}, the adiabatic (Born-Oppenheimer) approximation \cite{Giron:2020wpx}, and the bosonic algebraic approach \cite{Majarshin:2021hex}.

On the other hand, as far as the spectroscopy of the ground state $cc\bar c\bar c$ tetraquark is concerned, verily, there is difference between this work and some phenomenological approaches. For instance, the gap between two values of the $1^5S_2$ $cc\bar c\bar c$ state mass procured by the NR model of this work and complex scaling method of Ref. \cite{Wang:2022yes} is virtually 1 GeV, engendering that the $T_{\psi\psi}(6200)$ and $T_{\psi\psi}(6600)$ states are absent in the tetraquark framework of Ref. \cite{Wang:2022yes}. Accordingly, the enigma of the authentic $1S$-wave $cc\bar c\bar c$ energy region necessitates the further delving of experiments, which is crucial to the mass spectrum of the fully charmed tetraquark states as a whole.

\subsubsection{$2S$-, $3S$-, and $4S$-wave}\label{subsec422}

The spectroscopic properties of the radially excited $S$-wave $cc\bar c\bar c$ states are less investigated than the ground state $1S$-wave $cc\bar c\bar c$ states. Even so, there are still the disparities in the mass spectra of the $2S$-, $3S$-, and $4S$-wave fully charmed tetraquark states predicted by sundry phenomenological approaches. As Table \ref{T4c2} displays, the lowest and highest theoretical values of the $2S$-wave $cc\bar c\bar c$ states are 6480 \cite{Wang:2020ols} and 7281 MeV \cite{Wang:2022yes}, respectively. This work predicts that the masses of the $2S$-wave $cc\bar c\bar c$ states are in the scope between 6644 and 6864 MeV, in good agreement with the predictions of Refs. \cite{Wang:2022xja,Bedolla:2019zwg,Mutuk:2021hmi,Ke:2021iyh,Tiwari:2021tmz,Debastiani:2017msn,Lloyd:2003yc,Giron:2020wpx,Faustov:2021hjs,Jin:2020jfc,Wang:2019rdo,
Karliner:2020dta}. Furthermore, the $3S$-wave $cc\bar c\bar c$ states masses predicted by this work locate in the interval between 6979 and 7230 MeV, well conforming to the results of Refs. \cite{Wang:2021kfv,Wang:2019rdo,Ke:2021iyh,Wang:2022xja,Wang:2020ols,Bedolla:2019zwg,Tiwari:2021tmz,Lu:2020cns,Jin:2020jfc,Zhao:2020zjh}. Notwithstanding the predictions of the $4S$-wave $cc\bar c\bar c$ states are very few, the predicted results of this work, situated on the sector between 7213 and 7530 MeV, are in accordance with the most theoretical values \cite{Bedolla:2019zwg,Wang:2022xja,Wang:2020ols}.

\subsubsection{$1P$- and $2P$-wave}\label{subsec423}

As the first orbital excitations of the $S$-wave fully charmed tetraquarks, the $P$-wave $cc\bar c\bar c$ states are foremost perused by Ref. \cite{Chao:1980dv}. As unveiled in Table \ref{T4c3}, the spectroscopic properties of the $1P$-wave $cc\bar c\bar c$ states have been surveyed by a variety of phenomenological approaches \cite{Chao:1980dv,Tiwari:2021tmz,Debastiani:2017msn,Bedolla:2019zwg,Faustov:2021hjs,Ader:1981db,Giron:2020wpx,Deng:2020iqw,Badalian:1985es,Wang:2021kfv,liu:2020eha,
Lloyd:2003yc,Sonnenschein:2020nwn,Yu:2022lak}. Among them are the highest value 7110 MeV and lowest value 6390 MeV which are predicted by the holography inspired stringy hadron (HISH) model \cite{Sonnenschein:2020nwn} and quark bag model \cite{Chao:1980dv}, respectively. This work renders the spectroscopic results with the interval between 6478 and 6739 MeV as the potential masses of the $1P$-wave $cc\bar c\bar c$ states, congruent with the predictions from Refs. \cite{Chao:1980dv,Tiwari:2021tmz,Debastiani:2017msn,Bedolla:2019zwg,Faustov:2021hjs,Ader:1981db,Giron:2020wpx,Deng:2020iqw,Badalian:1985es,Wang:2021kfv}.

While not all of the formalisms accord with the results of this work well. One illustration is that the prediction on mass of the $1P$-wave $cc\bar c\bar c$ tetraquark in Ref. \cite{Sonnenschein:2020nwn} is overtly higher than the theoretical values in this work. However, Ref. \cite{Sonnenschein:2020nwn} accredits the possibility of not assigning the $T_{\psi\psi}(6900)$ state as the lowest ground state of the $cc\bar c\bar c$ tetraquark, which is capable of conducing to the $1P$-wave prediction not high any more. In addition, with regard to the masses of the $2P$-wave $cc\bar c\bar c$ states, a prediction with the energy range between 6867 and 7122 MeV is offered by this work, well coincident with the results of Refs. \cite{Tiwari:2021tmz,Bedolla:2019zwg,Debastiani:2017msn,Giron:2020wpx,Faustov:2021hjs}.

\subsubsection{$3P$-, $1D$-, $2D$-, and $1F$-wave}\label{subsec424}

Compared to the other low-lying excited $cc\bar c\bar c$ states, all of the $3P$-, $1D$-, $2D$-, and $1F$-wave $cc\bar c\bar c$ states are seldom probed by phenomenological theories, which incites the corresponding explorations of this work. It is apparent that the predictions on the mass spectra of the $3P$-, $2D$-, and $1F$-wave $cc\bar c\bar c$ states from the existing phenomenological theories are extremely exiguous \cite{Tiwari:2021tmz,Bedolla:2019zwg,Giron:2020wpx,Chao:1980dv,Sonnenschein:2020nwn}. In this work, the predicted values of the $3P$-, $1D$-, $2D$-, and $1F$-wave $cc\bar c\bar c$ states lie on the energy intervals between 7151 and 7432 MeV, between 6826 and 7002 MeV, between 7092 and 7325 MeV, and between 7035 and 7216 MeV, respectively. Manifestly, the results of the $1D$-wave fully charmed tetraquark states reaped by this work dovetail with the counterparts of Refs. \cite{Bedolla:2019zwg,Giron:2020wpx,Faustov:2021hjs,Liu:2021rtn,Deng:2020iqw}, as laid out in Table \ref{T4c4}.

\subsection{Expositions on the $T_{\psi\psi}$ states}\label{subsec43}

The potential tetraquark interpretations of the newly observed $T_{\psi\psi}(6200)$, $T_{\psi\psi}(6600)$, $T_{\psi\psi}(6900)$, and $T_{\psi\psi}(7300)$ states are displayed as follows.

\subsubsection{$T_{\psi\psi}(6200)$}\label{subsec431}

The $T_{\psi\psi}(6200)$ state, whose mass lay around 6220 MeV, was discovered by the ATLAS collaboration recently \cite{ATLAS:2022hhx}. Evidently, the mass $6220\pm50$ MeV of the $T_{\psi\psi}(6200)$ state is very close to the ground state mass 6200 MeV of the fully charmed tetraquark predicted by Ref. \cite{Iwasaki:1975pv}. Besides, the predicted value of the $1^5S_2$ $cc\bar c\bar c$ state mass in Table \ref{cccc1}, from the MGI ($\mu=100$) model in this work, is also close to the mass of the $T_{\psi\psi}(6200)$ state, which means that it is probable to assign the $T_{\psi\psi}(6200)$ state as the $1S$-wave fully charmed tetraquark state. Furthermore, numerous other predicted masses of the $1S$-wave $T_{\psi\psi}$ states, acquired by multifarious theories, lie on the energy domain between 6170 and 6270 MeV in unison, making the $1S$-wave interpretation of the $T_{\psi\psi}(6200)$ state more credible \cite{Bedolla:2019zwg,Berezhnoy:2011xn,Majarshin:2021hex,Wu:2016vtq,Barnea:2006sd,Deng:2020iqw,Zhuang:2021pci,Jin:2020jfc,Faustov:2020qfm,Karliner:2016zzc,
Nefediev:2021pww,Lundhammar:2020xvw,Iwasaki:1975pv,Wang:2022xja,Ke:2021iyh,Giron:2020wpx,Weng:2020jao}. Particularly, the masses $6200\pm100$ and $6270\pm90$ MeV of the $1^1S_0$ and $1^5S_2$ $cc\bar c\bar c$ states, predicted by the QCD sum rules approach, are in good agreement with the experimental value $6220\pm50$ MeV of the $T_{\psi\psi}(6200)$ state mass \cite{Wang:2022xja}.

On the other hand, as demonstrated in Table \ref{T4c1}, some theoretical values of the $cc\bar c\bar c$ ground state masses are lower or higher than the mass of the $T_{\psi\psi}(6200)$ state, varying within a large interval between 5300 and 7438 MeV \cite{Heupel:2012ua,Tiwari:2021tmz,Debastiani:2017msn,Wang:2017jtz,Wang:2018poa,Heller:1985cb,Zhu:2020xni,Wu:2022qwd,Anwar:2017toa,Karliner:2020dta,Pal:2023yid,
Ader:1981db,Mutuk:2021hmi,Gordillo:2020sgc,Chen:2021crg,Lloyd:2003yc,An:2022qpt,Zhang:2022qtp,Li:2021ygk,Yang:2021hrb,Wang:2019rdo,
Albuquerque:2020hio,Chen:2016jxd,Zhang:2020xtb,Zhao:2020zjh,Pal:2021gkr,Yang:2020atz,Zhao:2020nwy,Liu:2019zuc,Badalian:1985es,Wang:2021kfv,Lu:2020cns,
Asadi:2021ids,Sonnenschein:2020nwn,Esau:2019hqw,Wang:2022yes,Kuang:2022vdy}. Hence, the $T_{\psi\psi}(6200)$ state is awaiting the further experimental conformation. Additionally, it is worthy of noting that the decay width of the $T_{\psi\psi}(6200)$ state is extremely broad, indicating that the $T_{\psi\psi}(6200)$ state detected by the ATLAS collaboration is likely a short-lived resonance state.

\subsubsection{$T_{\psi\psi}(6600)$}\label{subsec432}

From the experimental points of view, two sorts of LHCb fit models, applied to the $T_{\psi\psi}(6600)$ data of the CMS collaboration, engendered a large difference around 200 MeV between the masses obtained by Fit Model I (labelled as the $T_{\psi\psi}(6600)_{\rm C}$ state) and Fit Model II (labelled as the $T_{\psi\psi}(6600)_{\rm C'}$ state) \cite{CMS:2022yhl}. Thereinto, the mass $6552\pm10$ MeV of the $T_{\psi\psi}(6600)_{\rm C}$ state is close to the mass $6620\pm30$ MeV of the state reported by the ATLAS collaboration (labelled as the $T_{\psi\psi}(6600)_{\rm A}$ state) \cite{ATLAS:2022hhx}, and the mass $6736\pm38$ MeV of the $T_{\psi\psi}(6600)_{\rm C'}$ state is very close to the mass $6741\pm6$ MeV of the state obtained by applying Fit Model II to the $T_{\psi\psi}(6600)$ data of the LHCb collaboration (labelled as the $T_{\psi\psi}(6600)_{\rm L'}$ state) \cite{LHCb:2020bwg}. All of the masses of the $T_{\psi\psi}(6600)$ states accord with the predicted masses of the $1P$- and $2S$-wave fully charmed tetraquark states from the GI, MGI ($\mu=50$), MGI ($\mu=100$), and NR models in this work. Consequently, the $T_{\psi\psi}(6600)$ states are good candidates of the $1P$- and $2S$-wave fully charmed tetraquark states.

Likewise, as Table \ref{T4c2} unveils, a number of theoretical masses of the $2S$-wave $cc\bar c\bar c$ states are close to the masses of the $T_{\psi\psi}(6600)$ states \cite{Wang:2020ols,Zhu:2020xni,Wang:2022xja,Bedolla:2019zwg,Mutuk:2021hmi,Tiwari:2021tmz,Debastiani:2017msn,Lloyd:2003yc,Giron:2020wpx}. Moreover, the experimental values of the $T_{\psi\psi}(6600)$ states masses are also consistent with the most predicted values of the $1P$-wave $cc\bar c\bar c$ states masses in Table \ref{T4c3} \cite{Chao:1980dv,Tiwari:2021tmz,Debastiani:2017msn,Bedolla:2019zwg,Faustov:2021hjs,Ader:1981db,Giron:2020wpx,Deng:2020iqw,Badalian:1985es,
Wang:2021kfv,liu:2020eha}. Therefore, the experimental determination on the parities of the $T_{\psi\psi}(6600)$ states is crucial to identify the nature of them in the future. On the other hand, there are several theoretical predictions that prefer to regard the $T_{\psi\psi}(6600)$ states as the $1S$-wave $cc\bar c\bar c$ ground states, making the further progress of experiments more necessary \cite{Nefediev:2021pww,Zhao:2020zjh,Lu:2020cns,Deng:2020iqw,Wu:2022qwd}.

\subsubsection{$T_{\psi\psi}(6900)$}\label{subsec433}

As the first $cc\bar c\bar c$ structure discovered by experiments, the $T_{\psi\psi}(6900)$ state is the $T_{\psi\psi}$ state which possesses the most experimental information in the entire members of the $T_{\psi\psi}$ family \cite{LHCb:2020bwg}. So far, as Table \ref{exp} reveals, the number of the $T_{\psi\psi}(6900)$ states reported by various experiments is the largest, compared to the other $cc\bar c\bar c$ states \cite{LHCb:2020bwg,ATLAS:2022hhx,CMS:2022yhl}. Although two sorts of LHCb fit models are also applied to the $T_{\psi\psi}(6900)$ data of the LHCb and CMS collaborations, the corresponding difference between the masses obtained by Fit Model I and Fit Model II is tiny. Not only that, the four masses $6905\pm11$, $6886\pm11$, $6927\pm9$, and $6918\pm10$ MeV of the $T_{\psi\psi}(6900)$ states observed by the LHCb and CMS collaborations are in accordance with the mass $6870\pm30$ MeV of the $T_{\psi\psi}(6900)$ state detected by the ATLAS collaboration in the $J/\psi$+$J/\psi$ channel (labelled as the $T_{\psi\psi}(6900)_{\rm A}$ state). Bizarrely, as far as the $T_{\psi\psi}(6900)$ states are concerned, a conundrum that the $2S$-wave assignment of the GI model is distinct with the $1D$- and $2P$-wave assignments of the other MGI ($\mu=50$), MGI ($\mu=100$), and NR models emerged in this work. By retrospecting the spectroscopic properties of heavy flavored mesons, the masses of highly excited states predicted by the GI model are regularly higher than the counterparts acquired by experiments and other theoretical models, e.g., the MGI and NR models \cite{Barnes:2005pb,Song:2015nia,Song:2015fha}. Thereupon, the $1D$- and $2P$-wave assignments from the MGI ($\mu=50$), MGI ($\mu=100$), and NR models are deemed as the most potential interpretations of the $T_{\psi\psi}(6900)$ states.

Additionally, the $1D$-wave assignment and $2P$-wave assignment of the $T_{\psi\psi}(6900)$ states are also endorsed by Refs. \cite{Giron:2020wpx,Faustov:2021hjs} and \cite{Tiwari:2021tmz,Debastiani:2017msn}, respectively. Wherefore, it is requisite to determine the parities of the $T_{\psi\psi}(6900)$ states by dint of experiments. On the other hand, the probability of the $3S$-wave assignment of the $T_{\psi\psi}(6900)$ states cannot be excluded in consideration of several theoretical predictions, which means that the experimental information of the $T_{\psi\psi}(6900)$ states is still insufficient \cite{Mutuk:2021hmi,Zhu:2020xni,Wang:2019rdo,Ke:2021iyh,Wang:2022xja}. Remarkably, the state reported by the ATLAS collaboration in the $J/\psi$+$\psi$(2S) channel (labelled as the $T_{\psi\psi}(6900)_{\rm A'}$ state), whose mass is $6780\pm360$ MeV, is left out in the spectroscopic comparison of experiments and theoretical models of this work, owing to the extremely broad interval of the mass value.

\subsubsection{$T_{\psi\psi}(7300)$}\label{subsec434}

Hitherto, there are two $T_{\psi\psi}(7300)$ states with the masses $7220\pm30$ and $7287\pm19$ MeV, which have been reported by the ATLAS and CMS collaborations \cite{ATLAS:2022hhx,CMS:2022yhl}, respectively. As mentioned previously, the assignments of the $T_{\psi\psi}(7300)$ states from the GI model are omitted due to the possible overestimation on the predicted masses of the highly excited states in heavy flavored mesons \cite{Barnes:2005pb,Song:2015nia,Song:2015fha}. On the basis of the predictions of the MGI ($\mu=50$), MGI ($\mu=100$), and NR models in Table \ref{cccc2}, the $T_{\psi\psi}(7300)$ states are able to be assigned as the $2D$-, $3P$-, or $4S$-wave fully charmed tetraquark states. Apart from that, as Tables \ref{T4c2} and \ref{T4c4} lay out, the $2D$-wave assignment, $3P$-wave assignment, and $4S$-wave assignment of the $T_{\psi\psi}(7300)$ states are also championed by Refs. \cite{Giron:2020wpx}, \cite{Tiwari:2021tmz,Bedolla:2019zwg}, and \cite{Bedolla:2019zwg,Wang:2022xja}, respectively. Thus, the parities of the $T_{\psi\psi}(7300)$ states are the pivotal hints to decipher the nature of them, awaiting the further delving of experiments.

\section{Summary}\label{sec5}

In the wake of a slew of experimental discoveries of heavy flavored exotic hadron states in recent decades, the inquiries on hadron spectroscopy are on the brick of a new era. In particular, the emergence of several fully charmed tetraquark states reported by the LHCb \cite{LHCb:2020bwg}, ATLAS \cite{ATLAS:2022hhx}, and CMS \cite{CMS:2022yhl} collaborations lately, promotes the further investigations on the family of heavy flavored multiquarks powerfully. Thereupon, in light of the diquark-antidiquark scenarios, this work aims to go in quest of the potential tetraquark interpretations of the newly observed $T_{\psi\psi}$ states systematically. Concretely, three sorts of diquark models, comprising the Godfrey-Isgur (GI) relativized diquark model, the modified Godfrey-Isgur (MGI) relativized diquark model with the color screening effects, and the nonrelativistic (NR) diquark model, are performed for the sake of endeavoring to unravel the nature of the $T_{\psi\psi}(6200)$, $T_{\psi\psi}(6600)$, $T_{\psi\psi}(6900)$, and $T_{\psi\psi}(7300)$ structures.

Overall, in this work, the $1S$-wave tetraquark assignment of the $T_{\psi\psi}(6200)$ state is perspicuously championed by the MGI diquark model. In the case of the $T_{\psi\psi}(6600)$ state, both of the $1P$- and $2S$-wave tetraquark assignments are prevalently endorsed by the GI, MGI, and NR diquark formalisms. With regard to the $T_{\psi\psi}(6900)$ state, both of the $1D$- and $2P$-wave tetraquark assignments are favored by the MGI and NR diquark approaches. In terms of the $T_{\psi\psi}(7300)$ state, all of the $2D$-, $3P$-, and $4S$-wave tetraquark assignments are the potential interpretations on the basis of the MGI and NR diquark scenarios. It is facile to be aware of that the experimental determination of parities is the fateful testimony to disentangle the nature of the newly observed $T_{\psi\psi}$ states. Ergo, the further experimental explorations on the $cc\bar c\bar c$ structures are obviously indispensable in the future, awaiting to be attained by the LHCb, ATLAS, CMS, and other collaborations.

\section*{ACKNOWLEDGMENTS}

W. C. Dong would like to thank Duojie Jia for helpful discussions and valuable comments. This work is supported by the National Natural Science Foundation of China (Grant No. 12175068).

\end{document}